\documentstyle[12pt]{article}
\def\ss{\scriptscriptstyle}
\def\Schrodinger{Schr\" odinger }
\newcommand{\phii}{\mbox{\boldmath $\phi$}}
\newcommand{\pii}{\mbox{\boldmath $\pi$}}
\begin{document}
\baselineskip=22pt plus 0.2pt minus 0.2pt
\lineskip=22pt plus 0.2pt minus 0.2pt
\font\bigbf=cmbx10 scaled\magstep3
\begin{center}
 {\bigbf Quantum Fields at Any Time}\\

\vspace*{0.35in}

\large

Charles G. Torre$^{1}$ and
Madhavan Varadarajan$^{2,3}$
\vspace*{0.25in}

\normalsize

{\sl
$^{1}$Department of Physics, Utah State University,}\\
{\sl Logan,
UT 84322-4415, USA}
\\
torre@cc.usu.edu\\
\medskip
{\sl $^{2}$Department of Physics, University of Utah,}\\
{\sl Salt Lake City,
UT 84112, USA}
\\
\medskip
{\sl $^{3}$Raman Research Institute,
Bangalore 560 080, India${}^*$}
\\
madhavan@rri.ernet.in\\
\vspace{.5in}
ABSTRACT

\end{center}
The canonical quantum theory of a free field using {\it
arbitrary foliations}
of a flat two-dimensional spacetime is investigated. It is
shown
that 
dynamical evolution along arbitrary spacelike foliations is 
unitarily implemented on the same Fock space as that
associated with
inertial foliations. It follows that the \Schrodinger
picture exists for arbitrary foliations
as a unitary image of the Heisenberg picture for the
theory. An explicit construction of the \Schrodinger
picture image of the Heisenberg Fock space states is
provided. The results presented here can be interpreted in
terms of
a Dirac constraint quantization of
parametrized field theory. In particular, it is shown
that 
the \Schrodinger picture physical states satisfy a functional
\Schrodinger 
equation which includes a slice-dependent $c$-number quantum
correction, in
accord with a proposal of Kucha\v r. 
The spatial diffeomorphism invariance of the \Schrodinger
picture
physical states is established.
Fundamental difficulties arise
when trying to generalize these results to
higher-dimensional
spacetimes.

\vfill\noindent
${}^*$ Permanent address.
\pagebreak

\setcounter{page}{1}

\section*{1. Introduction} 
The Poincar\'e invariant quantum theory of a free field is,
for all
practical purposes, completely understood \cite{Bogolubov,
Glimm, Haag}.
Most canonical
quantization treatments are in the Heisenberg 
picture and focus on the behavior of quantum fields
relative to
inertial foliations ({\it i.e.}, foliations by flat time
slices) of
the 
spacetime. In particular, the energy-momentum and angular
momentum of
the quantum field are densely defined self-adjoint
operators on a Fock space, which
generate unitary dynamical evolution {}from one flat
slice to
another. 

It is often assumed that the state of a quantum field in
flat
spacetime can be defined at {\it any} time, that is, upon an
arbitrary spacelike hypersurface. Likewise, it is assumed
that one can
define unitary dynamical evolution along an arbitrary
spacelike foliation of the
spacetime. While such niceties are apparently unnecessary
for a
non-gravitational treatment of particles and their
interactions, they become
interesting---if not mandatory---when trying to implement
some aspects of Einstein's
general theory of relativity in the quantum regime.
In this context there are no preferred foliations of
spacetime and
general covariance requires that all spacelike foliations
should be
allowed in the description of dynamics. 
Given the technical and
conceptual complexities that arise in attempts to construct
a quantum
theory of gravitation, it is useful to eliminate the
intricate
effects of the gravitational interaction and
focus on
the more limited --- but still non-trivial --- interplay
between quantum field theory and general covariance in
a flat spacetime. Thus it is of interest to examine free
quantum field theory in the context of an 
arbitrary spacelike foliation of the 
Minkowskian background. In this paper we focus our
attention
on two-dimensional
spacetimes since here the investigation can be completed
using
standard Fock space methods, and many of the mathematical
underpinnings for the investigation have already been
developed in \cite{GSegal}.  
Our primary concern is to establish whether operator
evolution from one arbitrary slice to another is unitarily
implemented  on the standard Fock space. If the evolution
is unitary, then  the most straightforward assignation of 
quantum states to slices is via the unitary image of the
states in the  (slice independent) Fock space. If unitarity
fails (as it seems to in dimensions higher than 2), it is an
open  question as to how one may assign states to slices. We
do not  address this question, other than hinting that the
algebraic approach  may be one way of addressing it.

Apart {}from the intrinsic
interest of
these issues {}from the
point of
view 
of quantum field theory on arbitrary foliations, 
this investigation can be viewed in terms of a
Dirac constraint quantization of parametrized scalar field
theory,
such as was considered by Kucha\v r \cite{kuchar1}. The
quantum parametrized field theory, being a
field theory possessing a diffeomorphism gauge group, is
often studied
as a model for some issues that arise in quantum gravity.
Indeed, in many ``midisuperspace'' models of general
relativity 
 one can identify the resulting reduced field theory with a 
parametrized field theory of one or more fields propagating
on a fixed
(often flat) spacetime (see, {\it e.g.}, \cite{TR}). 
Successful quantization of these 
models thus requires one to construct a suitable quantum
parametrized 
field theory.
In the usual approach to canonical quantization of such
diffeomorphism invariant 
field theories one aspires to use operator representatives
of the 
classical constraint functions to define a Hilbert space of
physical 
states.  The imposition of the 
quantum constraints is viewed as defining
unitary 
 transformations of states corresponding to evolution from
one 
(arbitrary) spacelike slice to another.  Even for the
parametrized 
theory of free fields 
propagating upon a two-dimensional spacetime it has been an
open
question whether such an approach can be rigorously
implemented.  We 
shall see that, in this case, the quantization can be
completed in the 
desired fashion.  On the other hand, it turns out that a 
straightforward generalization of these methods to
higher-dimensional 
models is not available.  Thus our investigation indicates
that 
alternative approaches ({\it e.g.}, 
algebraic approaches) to canonical quantization 
of generally covariant field theories become necessary
already in the 
simplest models for canonical quantum gravity.

A succinct
formulation of the problem addressed in this paper can be 
presented in the context of the algebraic formulation of
the quantization  of linear field theories on a fixed
background spacetime, which is by  now standard \cite{Haag,
Waldbook}.  The  $C^{*}$ algebra of observables is
traditionally taken to be  the Weyl algebra $\cal A$
associated with  the symplectic vector space of solutions
$\cal S$ to the field equations.   Quantum states  are
identified with positive linear functions on $\cal A$. 
Given  any pair of Cauchy surfaces $(\Sigma_{1},
\Sigma_{2})$, there is a symplectic  transformation 
$\tau\colon {\cal S} \to {\cal S}$ which can be interpreted
as  classical time  evolution from $\Sigma_{1}$ to
$\Sigma_{2}$.  This symplectic  transformation defines an
automorphism of $\cal A$ which is naturally  interpreted as
time evolution from $\Sigma_{1}$ to $\Sigma_{2}$ in  the
Heisenberg picture.  Now suppose that we associate a
state  $\omega_{1}\colon {\cal A}\to {\bf C}$
 (${\bf C}$ denotes the space of complex numbers)
  to the
instant of time represented by $\Sigma_{1}$.  (An
interesting,   potentially thorny issue is how one 
explicitly prepares/determines such a state on an arbitrary
slice.  We  hope to return to this question in future work.
)     By pull-back, the time evolution automorphism can be
viewed as  determining a new state,  $\omega_{2}$, which is
naturally interpreted as the Schrodinger  picture state at
the  instant of time 
defined by $\Sigma_{2}$.  A natural question that arises is
whether 
 this dynamical 
evolution can be expressed 
in terms of a unitary transformation on a Hilbert space 
representation of the Weyl algebra.  We will be considering
a free  field on Minkowski spacetime, so we focus on the
standard,  Poincar\' e invariant Fock  representation of
the Weyl algebra. Thus the question we wish to  address in
this paper is whether the automorphism of $\cal A$
associated  with a  pair of arbitrary Cauchy surfaces can
be realized as a unitary transformation  on the Fock space
representation of $\cal A$.   Because we are restricting
attention to free  fields, the investigation of this issue
can be given a completely  equivalent mathematical
formulation in terms  of unitary implementability of
dynamical evolution of operator valued  distributions
corresponding to Cauchy data (canonical coordinates and
momenta)  along an arbitrary foliation of spacetime by
Cauchy surfaces.  For free fields, the spatially smeared 
canonical coordinates and  momenta are observables in the
sense that  they are densely defined self-adjoint operators
on Fock space obtained  by a limiting procedure from the
Weyl observables.  We  must leave open the physical issues
regarding the sense in which the  quantum field on  an
arbitrary hypersurface is be interpreted, measured, {\it
etc.}  We  should also point out that there is no
compelling evidence to suggest that,  for Poincar\' e
invariant {\it interacting} field theories,   there
exist   observables corresponding to spatially smeared
Cauchy data. We prefer to formulate our investigation of
free field theory  in terms of canonical coordinates and 
momenta for a couple of reasons: (1) this is the 
formulation used in \cite{kuchar1}, whose results we are
trying to  extend; (2) in canonical quantum gravity,  for
which this work is intended as  a humble model, one
formulates the quantization problem in terms of 
``observables'' constructed from operator representatives
of  (functions of) Cauchy data for the field equations. 

Our investigation proceeds as follows.
Using the standard Fock space representation of a free
scalar field
on a
two-dimensional flat spacetime
we consider Heisenberg picture field operators
(operator-valued
distributions)
associated with arbitrary (curved) spacelike slices.
We ask whether the evolution of
field
operators
{}from one such slice to another, as dictated by the field
equations,
is unitarily implemented on
the Fock space.
This issue, although formulated in the context of
slice-dependent
operators 
in the Heisenberg picture, is intimately connected with the
existence
of the 
\Schrodinger picture. In the \Schrodinger picture, field
operators are 
slice-independent and are associated with some fixed
initial slice of
the 
foliation. The dynamics are encoded in the slice-dependent
state vectors
which, presumably, satisfy a functional \Schrodinger
equation,
usually associated with the names Tomonaga and Schwinger
\cite{Tomonaga, Schwinger}; see also the book of Dirac
\cite{Dirac}. Given a
foliation, if there exists a one-parameter 
family of unitary transformations which implement the
operator
evolution 
{}from slice to slice of the foliation, then the \Schrodinger
picture is 
defined as the unitary image of the Heisenberg picture. In
this paper
we
show that such unitary transformations exist for 
a free,
massless
scalar field propagating on a flat spacetime with manifold
structure $R \times S^1$, and we
investigate properties of the \Schrodinger picture quantum
states.  We thus largely complete the quantization
program initiated in \cite{kuchar1} by rigorously 
constructing the physical quantum states in the \Schrodinger
picture.
In so doing, we derive the anomaly
potential, proposed in \cite{kuchar1}, which appears in the
quantum
constraint equations as a $c$-number quantum correction.
With a
rigorous construction of the physical states in hand, it is
now
possible to investigate in detail various diffeomorphism
invariance-related 
issues in quantum field theory. In this paper we answer the
question:
to what extent are the physical states of the parametrized
quantum field theory
actually invariant under spatial diffeomorphisms? This
invariance is usually
assumed in approaches to canonical quantization of
diffeomorphism invariant field theories, but at least for
the two-dimensional models
such as considered here, spatial diffeomorphism
invariance 
is called into question by the
quantum corrections which appear in the constraints.

Let us emphasize what we are {\it not}
doing in this paper. We are not considering the effect of
classical
gravitational fields on quantum matter fields, which is the
subject
of quantum field theory in curved spacetime. We are not
considering
different quantization schemes in flat spacetime. The
complex
structure and Fock space that we use are the 
standard ones associated with the timelike Killing
vector field of
the
Minkowski metric and are 
fixed once and for all. So, for example, in this paper we
do not (explicitly) consider 
slice-dependent complex structures and Fock
spaces.
As mentioned before, the  simplest definition of
slice-dependent state is as the unitary image of a
Heisenberg picture state. We do not discuss how to 
measure/prepare such a state. We hope to return to this
question in a future work. Finally, we do not investigate
the feasability or existence of other definitions of slice
dependent states.

The outline of the paper is as follows. In \S2 we 
summarize 
the classical theory of a free scalar field on $R \times
S^1$, and we remind the reader of the standard 
Fock space quantization of the theory in the Heisenberg
picture. We provide the relation 
to the framework of parametrized field theory and its Dirac
quantization
as constructed in \cite{kuchar1}. Finally, we demonstrate
the existence 
of the unitary transformation which dictates evolution of
operators
{}from 
one time slice to another. 
In \S3, we construct the \Schrodinger
picture for the theory and give an
explicit construction of the \Schrodinger picture states on
an
arbitrary time slice as unitary
images of the Heisenberg states. We show that the
\Schrodinger picture states satisfy a functional
\Schrodinger equation which includes an embedding-dependent
quantum
correction relative to the classical equation. This
$c$-number correction is related to the ``anomaly
potential'' of
\cite{kuchar1}. 
Section 4 is devoted to the issue of spatial diffeomorphism
invariance of the 
solutions to the functional \Schrodinger equations. There we
relate
the factor ordering of the
spatial projection of the \Schrodinger equation
to a version of the Schwarzian derivative due to
Segal \cite{GSegal}.
This leads to an interpretation of the spatially covariant
``gauge''
choice advocated by Kucha\v r for the anomaly 
potential.   With this result in hand we are able to show
that the
functional \Schrodinger equation implies spatial
diffeomorphism
invariance of physical states in the \Schrodinger
representation.
In \S 5 we briefly consider generalizations of our results
to massive
free fields and to spacetimes with topology
$R^2$. We also indicate the fundamental difficulties
inherent in generalizing
our results to higher 
spacetime dimensions. \\

{\bf Notation}
Classical fields are distinguished {}from their quantum
counterparts
by 
adopting bold face type for the former ({\it e.g.}, $\phi
(x)$ is
the
quantum 
counterpart of the classical field $\phii(x)$).
Inertial 
coordinates on $R\times S^1$ are $T \in (-\infty,
\infty)$
and 
$X \in [0, 2\pi]$, with respect to which the line element
is 
\begin{equation}
ds^2 =-dT^2 + dX^2.
\end{equation}
We denote by $T^{\pm}:= T{\pm}X$ the 
advanced and retarded null coordinates. Derivatives with
respect to 
$T^{\pm}$ are denoted with the subscripts `$,\pm$' 
({\it e.g.},
$\phi_{,+}
=
{\partial\phi \over \partial T^+}$).
On a generic spacelike foliation
we denote the spatial coordinate on a leaf of the foliation
by $x\in
[0,2\pi]$. Spatial derivatives (with respect to $x$)
are denoted with the subscript 
``$,x$'' ({\it e.g.},
$f_{,x}(x)={df(x)\over dx}$).
Leaves of the foliation are labeled by the parameter
$t$. We
define a foliation by specifying the parametric equations
\begin{equation}
T^{\alpha}=T^{\alpha}(t,x),
\end{equation}
where the superscript $\alpha$ labels coordinates on 
$R\times S^1$, 
{\it e.g.}, $T^\alpha=(T,X)$ or $T^\alpha=(T^+,T^-)$, and
\begin{equation}
T^+ _{,x}(t,x)>0,\quad T^-_{,x}(t,x)<0,
\label{eq:ineq}
\end{equation}
\begin{equation}
T^\pm(t,2\pi)=T^\pm(t,0)\pm 2\pi.
\label{eq:discont}
\end{equation}
A particular spacelike slice is determined by
an {\it embedding}: 
\begin{equation}
T^\alpha=T^\alpha(x),
\end{equation}
which can be identified with a leaf $t=t_0$ of a
foliation via
$$
T^\alpha(x)=T^\alpha(t_0,x).
$$

\section*{2. The Heisenberg picture for a free massless
scalar field
on $R \times S^1$}
\subsection*{2a. The classical theory}

The massless scalar field on $R \times S^1$ satisfies
the
wave equation
\begin{eqnarray}
\Box \phii & = & 0 , \label{eq:KG}\\
\Rightarrow  \phii (T^+, T^-) & = & 
{\phii }^{+}(T^+)
\;
 + \; { \phii }^{-}(T^-) . \label{eq:pm}
\end{eqnarray}
We expand the scalar field in modes as
\begin{equation}
{ \phii^{\pm}}= 
 {1 \over \sqrt{2\pi}}\Bigg[ {1\over 2}({\bf q}+ {\bf
p}T^{\pm})
 +{1\over
\sqrt{2}}\sum_{k=1}^{\infty} 
 ({1\over \sqrt{k}}{\bf
a}_{(\pm)k}e^{-ikT^{\pm}}+
{1\over \sqrt{k}}{\bf a}^{*}_{(\pm)k}e^{ikT^{\pm}}) \Bigg].
\label{eq:modexp}
\end{equation}
The real numbers ${\bf q, p}$ will be referred to as the
zero modes 
of the field. The complex numbers ${\bf a}_{(+)k},{\bf
a}_{(-)k}$ 
and their complex conjugates ${\bf a}^{*}_{(+)k},{\bf
a}^{*}_{(-)k}$ 
are the familiar Fourier mode
coefficients (note that $k>0$).

The field can be restricted to an embedding ({\it i.e.}, a
leaf of
a
foliation)
$T^\alpha=T^\alpha(x)$, which results in
the
definition
\begin{equation}
{ \phii}(x) :={\phii} (T^\alpha(x)) 
= {\phii}^{+}(T^+(x))
\;
+ \; {\phii}^{-}(T^-(x)).
\label{eq:phi}
\end{equation}
Given an embedding $T^\alpha(x)$, we also define 
\begin{equation}
\pii (x) :=
\sqrt{\gamma}n^\alpha\nabla_\alpha\phii
\Big|_{T^\alpha=T^\alpha(x)},
\end{equation}
where $\sqrt {\gamma}$ is the
determinant of
the 
1-metric induced on the spatial slice and $n^\alpha$ is the 
future-pointing unit normal to the slice. Thus $\pii (x)$ is
the field
momentum associated with the given embedding.
A simple computation shows that
\begin{equation}
{\pii} (x) = T^{+}_{,x}(x) {\phii}_{,+}(T^+(x)) 
 - T^{-}_{,x}(x){\phii}_{,-}(T^-(x)) .
\label{eq:pi}
\end{equation}

The slice-dependent fields $(\phii (x),\pii (x))$ are Cauchy
data for (\ref{eq:KG}) and provide a canonical coordinate
chart on the phase space of solutions of the wave
equation. The wave equation can be used to determine the
evolution of the
fields 
$(\phii (x),  \pii (x))$ {}from one arbitrary slice
to
another. This
evolution is encoded in the following functional evolution 
equations :
\begin{eqnarray}
{\delta {\phii } (x) \over \delta T^{\pm}(x^{\prime})}
 & = & 
\pm {({\pii }(x)\pm {\phii }_{,x}(x)) \over
2T^{\pm}_{,x}(x)} 
 \delta (x,x^{\prime}), 
\label{eq:evolphi} \\
{\delta \pii (x) \over \delta T^{\pm}(x^{\prime})}
 & = & 
{({\pii }(x^{\prime})\pm {\phii }_{,x}(x^{\prime}))
 \over 2T^{\pm}_{,x}(x^{\prime})} 
 { \partial\delta
(x,x^{\prime})\over \partial
x} \; . 
\label{eq:evolpi}
\end{eqnarray}
In the context of a particular foliation,
$T^\alpha=T^\alpha(t,x)$, equations
(\ref{eq:evolphi}), 
(\ref{eq:evolpi})
give the infinitesimal change of $({ \phii } (x), {\pii }
(x))$
 corresponding to evolution {}from the slice $T^\alpha(x,t)$
to the slice
$T^\alpha(x,t+ dt)$ via 
\begin{equation}
{\partial \phii (x,t)\over\partial
t}=\int_0^{2\pi}{\partial 
T^\alpha(x^\prime,t)\over
\partial t}\,{\delta \phii (x,t)\over\delta
T^\alpha(x^\prime,t)}\, dx^\prime,
\label{eq:phi-dot}
\end{equation}
\begin{equation}
{\partial \pii (x,t)\over\partial t}=\int_0^{2\pi}{\partial 
T^\alpha(x^\prime,t)\over
\partial t}\,{\delta \pii (x,t)\over\delta
T^\alpha(x^\prime,t)}\, dx^\prime,
\label{eq:pi-dot}
\end{equation}
This time evolution is a one-parameter family of canonical
transformations which we would like to carry over into
unitary transformations in the quantum theory. In
particular, we 
shall deal with dynamical
evolution along an arbitrary foliation connecting a fixed
initial slice
$T^\alpha_0(x)$ to a slice $T^\alpha(x)$. Data on
$T^\alpha_0(x)$
will be denoted by 
$({ \phii}_{0} (x), {\pii}_{0} (x))$. For simplicity,
we
restrict attention
to the case where the initial slice of our foliation is
flat,
and
corresponds
to $T=0$ with arc-length parametrization. Thus 
\begin{equation}
T^{+}_{0}(x)= -T^{-}_{0}(x)=x,
\label{eq:initial}
\end{equation}
and 
$({ \phii }_{0} (x), { \pii }_{0} (x))$ are the equations 
(\ref{eq:phi}), (\ref{eq:pi}) evaluated on 
$T^\alpha_{0}(x)$. Equations 
(\ref{eq:evolphi}), (\ref{eq:evolpi}) with initial data 
$({ \phii }_{0} (x), {\pii }_{0} (x))$ on the initial
slice given
by 
(\ref{eq:initial}) can be solved to give a unique solution
to
(\ref{eq:KG}).

\subsection*{2b. Quantum theory: The Hilbert space}
We now consider the operators $q, p,
a_{(\pm)k},a^{\dagger}_{(\pm)k} $ 
corresponding to the classical 
quantities ${\bf q},{\bf p},{\bf a}_{(\pm)k},{\bf
a}^{*}_{(\pm)k} $. 
We recall
the standard Hilbert space construction \cite{kuchar1} on
which 
the only nontrivial commutation relations are
\begin{equation}
[q,p] = i{\cal I} \; , \label{eq:qp} 
\end{equation}
\begin{equation}
[ a_{(\pm)k}, a^{\dagger}_{(\pm)l}] = \delta_{kl}{\cal I},
\label{eq:fock}
\end{equation}
where $\cal I$ is the identity.  
The Hilbert space ${\cal H}$ of the theory is a product of
three
Hilbert spaces,
\begin{equation}
{\cal H} = {\cal F}^{(+)} \otimes {\cal F}^{(-)}\otimes
{\cal L}^2(R).
\label{eq:Hilbert}
\end{equation}
where ${\cal F}^{(\pm)}$ are the standard Fock spaces on
which the 
$a^\dagger_{(\pm)k},a_{(\pm)k}$ operators are
represented as creation and annihilation operators. 
${\cal L}^{2}(R)$ is the representation space for the zero
mode
operators $(q,p)$.

To illustrate our notation and conventions we recall the
standard
construction
 of the Fock space associated with the `+' operators. 
The vacuum state $|(+);0\rangle \in {\cal
F}^{(+)}$ is
such that 
\begin{equation}
a_{(+)k} |(+);0\rangle = 0\qquad \forall\, k .
\label{eq:annihilation}
\end{equation}
The normalized $N$-particle states are generated {}from 
$|(+);0\rangle $ by the action of the creation operators so
that 
\begin{equation}
|(+);n_{k_1}...n_{k_m}\rangle 
 := {(a^{\dagger }_{(+)k_1})^{n_{k_1}} \over
\sqrt{n_{k_1}!}}
...{(a^{\dagger }_{(+)k_m})^{n_{k_m}} \over
\sqrt{n_{k_m}!}} |0\rangle\;, 
\qquad\sum_{i=1}^{m}n_{k_i} =N .
\label{eq:creation}
\end{equation}
The vectors $|(+);n_{k_1},\dots,n_{k_m}\rangle \forall m,
\forall \{ k_i, n_{k_i},
i=1,\dots,m \} $
with $|(+);0\rangle $ 
form an orthonormal basis for ${\cal F}^{(+)}$. The action
of 
$a_{(+)k}$ on any state in this basis is obtained {}from 
(\ref{eq:fock}), (\ref{eq:annihilation}),
(\ref{eq:creation})
. 

The operators $a_{(-)k}, a^{\dagger}_{(-)k}$ are
represented in an 
identical manner on ${\cal F}^{(-)}$, while $q, p$ are
densely defined on 
${\cal L}^2 (R)$ in the usual way. For
our purposes, we find the momentum representation
convenient:
$p\psi ({\bf p})= {\bf p}\psi ({\bf p})$ and 
$q\psi ({\bf p})=i {d\psi \over d{\bf p}}$.

We identify the operator-valued distributions corresponding
to
(\ref{eq:phi}), (\ref{eq:pi}) by replacing 
${\bf p},{\bf q},{\bf a}_{(\pm)k},{\bf a}^{*}_{(\pm)k} $ in
these
expressions with the operators
$q, p, a_{(\pm)k},a^{\dagger}_{(\pm)k} $. Since the
classical
evolution 
equations are linear, the operator valued distributions
$\phi (x)$
and $\pi (x)$ satisfy the corresponding evolution equations
for
operators in the Heisenberg picture.
In \S2d we will show that the corresponding dynamical
evolution is unitarily implemented.

\subsection*{2c. Relation to parametrized field theory and
its Dirac
quantization}
It is a simple matter to check that the quantum system
described
above is
the same as that arising in 
the Heisenberg picture constraint
quantization of parametrized field theory developed in
\cite{kuchar1}.
The only
differences lie in our notation and different
normalizations for the
quantities (${\bf a}_{(\pm)k},{\bf a}^{*}_{(\pm)k} $ )
and their quantum counterparts. We briefly summarize the
treatment
of \cite{kuchar1} in our slightly 
different notation and conventions.

The phase space of
a parametrized,
free, massless, scalar field on the Minkowskian cylinder
consists
of the embedding fields $T^{\alpha}(x)$, and their conjugate
momenta $P_{\alpha}(x)$\footnote{The notation for the
classical embedding coordinates and
their
conjugate momenta is an exception to our convention of
denoting
classical
quantities by bold face type. This is to minimize confusion
with the
notation of \cite{kuchar1} in which bold face type does not
have the
same meaning as in this paper.}, 
along with the scalar field ${\phii} (x)$ and its
conjugate
momentum ${\pii} (x)$. Corresponding to the
diffeomorphism
invariance
of the parametrized theory, there are two
constraints
\begin{equation}
C_{\pm} = P_{\pm} \pm {({\pii}(x)\pm 
{\phii}_{,x}(x))^2
 \over 4T^{\pm}_{,x}(x)}\approx0,
\label{eq:constraint}
\end{equation}
which completely fix the embedding momenta in terms of the
remaining fields. These constraints are first class (they
have
strongly vanishing Poisson brackets) and indicate that the
embeddings can be viewed as ``pure gauge''. The
phase space
variables can be mapped via an embedding-dependent canonical
transformation
to a new set of phase space coordinates 
$({\bf P}_{\pm}(x), T^{\pm}(x),
{\bf p},{\bf q},{\bf a}_{(\pm)k},
{\bf a}^{*}_{(\pm)k}) $
via
(\ref{eq:modexp}--\ref{eq:pi}) \cite{kuchar1}. The
transformation
leaves the embedding fields unchanged, while
the new embedding momenta are the constraint functions:
\begin{equation}
{\bf P}_{\pm}(x) := C_{\pm}\approx0 .
\label{eq:hbergmom}
\end{equation}
This transformation hinges upon the fact that the constraint
functions $C_\alpha$ satisfy an Abelian Poisson algebra.
In these ``Heisenberg'' variables, the constraints are
therefore simply the vanishing of the embedding momenta.

Based upon the Heisenberg variables just described, 
Kucha\v r
implements the Dirac constraint quantization of the
parametrized
field theory in the Heisenberg picture as follows.  In the
quantum
theory the operators
 $q, p, a_{(\pm)k},a^{\dagger}_{(\pm)k} $ are represented
as in
\S2b.
The embedding fields act by multiplication and the
embedding momenta act by functional differentiation.
The quantum constraints,
\begin{equation}
{\bf P}_\alpha |\Psi> = {1\over i}{\delta\over\delta
T^\alpha} |\Psi>
= 0, 
\end{equation}
then imply that the physical states
are time independent, that is, independent of the embedding.
The physical states can thus be identified with the
embedding-independent Fock states of \S2b. Thus, constraint
quantization based upon the
canonical variables $({\bf P}_{\pm}(x), T^{\pm}(x),
{\bf p},{\bf q},{\bf a}_{(\pm)k},
{\bf a}^{*}_{(\pm)k}) $, corresponds exactly to the
canonical quantum theory in the Heisenberg picture outlined
in
\S2b.  

{}From the point of view of Dirac quantization of
parametrized field
theory, our primary goal in this paper is to recover 
the quantum theory
in the
\Schrodinger picture.  In particular, we aim to obtain
physical states
satisfying quantum constraints of the form
\begin{equation}
\widehat C_\pm |\Psi> = 0,
\end{equation}
where 
$\widehat C_\pm$ is a quantum version of the classical
constraint
function (\ref{eq:constraint}).
\subsection*{2d. Unitarity of time evolution}

For each embedding, the quantum fields $(\phi(x), \pi(x))$
generate a
*-algebra of observables via their canonical commutation
relations \cite{Haag}. In this section we show
that the observable algebras associated with different,
arbitrary time slices are unitarily equivalent. We do this
by comparing 
$(\phi(x),\pi(x))$ and $(\phi_0(x),\pi_0(x))$ and building
up the unitary transformation relating these operator-valued
distributions on each of ${\cal F}^{(+)}$,
${\cal F}^{(-)}$ and ${\cal
L}^{2}(R)$. To this end, expand the fields
$(\phi(x),\pi(x))$ and $(\phi_0(x),\pi_0(x))$ in Fourier
series:
\begin{equation}
\phi_0(x)={1\over\sqrt{2\pi}}\left(
q+{1\over\sqrt{2}}\sum_{k=1}^\infty {1\over\sqrt{k}}[
(a_{(+)k}+a^\dagger_{(-)k})e^{-ikx} +
(a_{(+)k}^\dagger+a_{(-)k})e^{ikx}]\right),
\label{eq:phi0}
\end{equation}
\begin{equation}
\pi_0(x)={1\over\sqrt{2\pi}}\left(
p-{i\over\sqrt{2}}\sum_{k=1}^\infty \sqrt{k}[
(a_{(+)k}-a^\dagger_{(-)k})e^{-ikx} -
(a_{(+)k}^\dagger-a_{(-)k})e^{ikx}]\right),
\label{eq:pi0}
\end{equation}
\begin{equation}
\phi(x)={1\over\sqrt{2\pi}}\left(
q[T]+{1\over\sqrt{2}}\sum_{k=1}^\infty {1\over\sqrt{k}}[
(a_{(+)k}[T]+a^\dagger_{(-)k}[T])e^{-ikx} +
(a_{(+)k}^\dagger[T]+a_{(-)k}[T])e^{ikx}]\right),
\label{eq:Heisphi}
\end{equation}
\begin{equation}
\pi(x)={1\over\sqrt{2\pi}}\left(
p[T]-{i\over\sqrt{2}}\sum_{k=1}^\infty \sqrt{k}[
(a_{(+)k}[T]-a^\dagger_{(-)k}[T])e^{-ikx} -
(a_{(+)k}^\dagger[T]-a_{(-)k}[T])
e^{ikx}]\right),
\label{eq:Heispi}
\end{equation}
where
\begin{equation}
a_{(\pm) k}[T] ={1\over2\pi\sqrt{k}}\int_{0}^{2\pi}
e^{\pm ikx} T^{\pm}_{,x}
\left[ \pm {ip \over \sqrt{2}} \pm 
\sum_{n=1}^\infty\sqrt{n}(a_{(\pm) n}e^{-inT^\pm(x)}-
a_{(\pm) n}^\dagger e^{ inT^\pm(x)})\right]\, dx,
\label{eq:Heisa}
\end{equation}
\begin{eqnarray}
q[T] 
 = 
q+{1\over2\pi}p\int_0^{2\pi}T(x)\, dx
 &+&
{1\over2\pi}\int_0^{2\pi}dx\Big({1\over \sqrt{2}}\sum_{n=
1}^\infty
{1\over\sqrt{n}}
\big[a_{(+)n}e^{-inT^+(x)}+a_{(+)n}^\dagger
e^{inT^+(x)\nonumber}\\
 & &+a_{(-)n}e^{-inT^-(x)}
+a^\dagger_{(-)n}
e^{inT^-(x)}\big]\Big)\, dx,
\label{eq:Heisq}
\end{eqnarray}
\begin{equation}
p[T]=p .
\label{eq:Heisp}
\end{equation}

It is straightforward to verify at a purely algebraic
level (that is, ignoring issues of domain), that the
commutation relations between the variables 
(\ref{eq:Heisphi}), (\ref{eq:Heispi}) are independent of
the embedding fields
$T^\pm(x)$. In other words, $(q[T],p[T],
a_{(\pm) k}[T],a^\dagger_{(\pm) k}[T])$ have the
non-vanishing commutators given in (\ref{eq:qp}),
(\ref{eq:fock}). The transformation 
\begin{equation}
(q[T],p[T],
a_{(\pm) k}[T],a^\dagger_{(\pm)
k}[T])\longleftrightarrow(q,p,
a_{(\pm) k},a^\dagger_{(\pm) k})
\label{eq:symplectic}
\end{equation}
is a symplectic transformation which is a quantum
analog of the canonical transformation mentioned in \S2c. 
We now want to see that there is an embedding-dependent
unitary transformation $U=U[T]$ on ${\cal H}$ such that
\begin{equation}
q[T]=U^\dagger q U,\qquad p = U^\dagger p U,\qquad 
a_{(\pm) k}[T]=U^\dagger a_{(\pm) k} U.
\end{equation}
The basic theory of the unitary implementability on Fock
space of
symplectic transformations on the vector space 
of solutions to linear field equations
is due to Shale \cite{Shale}, see also \cite{Wald}. Because
of the
existence of the zero modes, we find it convenient to first
decompose the symplectic transformation
(\ref{eq:symplectic}) into two successive symplectic
transformations, and then check that each transformation is
unitarily implementable. To this end, we view the
transformation 
(\ref{eq:symplectic}) as being defined by the composition
of the symplectic transformation

{\bf (I)}
\begin{equation}
(q,p,a_{(\pm) k},a^\dagger_{(\pm) k})
\longrightarrow (q,p,c_{(\pm) k}[T],
c^\dagger_{(\pm) k}[T]),
\end{equation}
where
\begin{equation}
c_{(\pm) k}[T] = \pm{1\over2\pi\sqrt{k}}\int_{0}^{2\pi}
e^{\pm ikx} T^{\pm}_{,x}
\left[\sum_{n=1}^\infty\sqrt{n}(a_{(\pm) n}e^{-
inT^\pm(x)}-
a_{(\pm) n}^\dagger e^{inT^\pm(x)})\right]\, dx,
\label{eq:c}
\end{equation}
followed by the symplectic transformation

{\bf (II)}
\begin{equation}
(q,p,c_{(\pm) k}[T],c^\dagger_{(\pm) k}[T])
\longrightarrow (q[T],p,a_{(\pm) k}[T],
a^\dagger_{(\pm) k}[T]),
\end{equation}
where
\begin{equation}
a_{(\pm) k}[T] =c_{(\pm)k}[T] \pm {ip \over \sqrt{2}}
{1\over2\pi\sqrt{k}}\int_{0}^{2\pi}\, e^{\pm ikx}\,
T^{\pm}_{,x}\,dx , 
\end{equation}
and $q[T]$ is defined in (\ref{eq:Heisq}).

Because $T^+(x)$ and $T^-(x)$ each define diffeomorphisms
of the circle (see 
(\ref{eq:ineq}), (\ref{eq:discont})), the transformation
{\bf (I)} involves two copies of the ``metaplectic
representation'' of the group Diff$(S^1)$, which is
discussed in \cite{GSegal}. It follows that the
transformation {\bf (I)}, for each sign $+$ and $-$, arises
as a unitary transformation
$U^{(\pm)}_{\rm\scriptscriptstyle I}[T]$ on ${\cal
F^{(\pm)}}$ (and the identity on the
zero mode sector of the Hilbert space):
\begin{equation}
U_{\rm\scriptscriptstyle I}^{(\pm)\dagger} q 
U_{\rm\scriptscriptstyle I}^{(\pm)} = q
\label{eq:UIq}
\end{equation}
\begin{equation}
U_{\rm\scriptscriptstyle I}^{(\pm)\dagger} p 
U_{\rm\scriptscriptstyle I}^{(\pm)} = p
\label{eq:UIp}
\end{equation}
\begin{equation}
U_{\rm\scriptscriptstyle I}^{(\pm)\dagger} a_{(\pm)k} 
U_{\rm\scriptscriptstyle I}^{(\pm)} = c_{(\pm)k}[T].
\label{eq:UIa}
\end{equation}
The gist of the proof involves showing that the Bogolubov
coefficients
\begin{equation}
B^{(\pm)}_{mn}[T] = 
\mp {1\over2\pi}\sqrt{{n\over m}}\int_{0}^{2\pi}e^{\pm
imx}
T^{\pm}_{,x}(x)
e^{ inT^\pm(x)}\, dx,
\end{equation}
are Hilbert-Schmidt, {\it i.e.}, satisfy
\begin{equation}
\sum_{m,n=1}^\infty|B^{(\pm)}_{mn}|^2 < \infty.
\label{eq:HS}
\end{equation}
This latter result is guaranteed if the embedding is taken
to be sufficiently smooth (see the Appendix). 

Next, it is straightforward to check that both
\begin{equation}
Z^{(\pm)}_n:={1\over2\pi\sqrt{n}}\int_0^{2\pi}e^{\mp
inx}T^{\pm}_{,x}dx
\end{equation}
and
\begin{equation}
\zeta_{(\pm)
n} := {1\over2\pi\sqrt{n}}\int_0^{2\pi}\,e^{inT^\pm(x)}\,
dx
\label{eq:zeta}
\end{equation}
are rapidly decreasing functions of $n$, that is, as 
$n\rightarrow \infty$, $|Z^{(\pm)}_n|$ and
$|\zeta^{(\pm)}_n|$ 
vanish faster
than any power of $1/n$. For details, see the Appendix. 
{}From this it follows that 
$U_{\rm\scriptscriptstyle II}[T]$, defined as 
\begin{eqnarray}
U_{\rm\scriptscriptstyle II}[T] &=& \exp\Big\{-i
\Big[{p^2\over4\pi}
\int_0^{2\pi}T(x) dx
-
\Big({p\over \sqrt{2}}\sum_{n=1}^\infty
\big[c_{(+)n}Z_n^{(+)}\nonumber\\ 
& &\qquad+c_{(+)n}^\dagger Z_n^{(+)*}
-c_{(-)n}Z_n^{(-)}
-c^\dagger_{(-)n}
Z_n^{(-)*}\big]\Big)\Big]\Big\}\nonumber \\
&=& \exp\Big\{-i
\Big[{p^2\over4\pi}
\int_0^{2\pi}T(x) dx
+
\Big({p\over\sqrt{2}}\sum_{n=1}^\infty
\big[a_{(+)n}\zeta_n^{(+)*}\nonumber\\ 
& &\qquad+a_{(+)n}^\dagger \zeta_n^{(+)}
+a_{(-)n}\zeta_n^{(-)*}
+a^\dagger_{(-)n}
\zeta_n^{(-)}\big]\Big)\Big]\Big\} ,
\label{eq:UII}
\end{eqnarray}
is a unitary operator on the Hilbert space ${\cal H}$.
$U_{\rm\scriptscriptstyle II}$ implements the
transformation {\bf (II)}:
\begin{equation}
U_{\rm\scriptscriptstyle II}^\dagger
qU_{\rm\scriptscriptstyle II} = q[T]
\end{equation}
\begin{equation}
U_{\rm\scriptscriptstyle II}^\dagger
pU_{\rm\scriptscriptstyle II} = p
\end{equation}
\begin{equation}
U_{\rm\scriptscriptstyle II}^\dagger c_{(\pm)
k}[T]U_{\rm\scriptscriptstyle II} = a_{(\pm)k}[T].
\end{equation}

The combined transformation $U[T]=
U^{(+)}_{\rm\scriptscriptstyle
I}U^{(-)}_{\rm\scriptscriptstyle I} 
U_{\rm\scriptscriptstyle II}$ is the unitary map
implementing dynamical evolution {}from the initial
spacelike
embedding $T_0^\pm(x)=\pm x$ to the final spacelike
embedding $ T^\alpha(x)=(T^+(x),T^-(x))$.

\section*{3. The \Schrodinger picture}

\subsection*{3a. \Schrodinger picture image of the Fock
basis}

A vector in the Hilbert space for the quantum field theory
is any normalizable superposition of the Fock basis vectors
(see {\S2b}). In the Heisenberg picture of dynamics, any
such vector 
can represent the state vector $|\Psi>_{\rm
\scriptscriptstyle H}$ of
the
system for all time. Dynamical results depend upon
specification of
an embedding, and are expressed
in terms of expectation values of observables built {}from
the embedding-dependent operator-valued distributions 
$(\phi(x), \pi(x))$ defined in \S2d. In the \Schrodinger
picture,
dynamical
evolution is encoded in embedding-dependent state vectors
$|\Psi[T]>_{\rm\scriptscriptstyle S}$ according to the
unitary mapping
\begin{equation}
|\Psi[T]>_{\rm\scriptscriptstyle
S}=U[T]|\Psi>_{\rm\scriptscriptstyle
H},
\end{equation}
and dynamical results are expressed in terms of operator
observables constructed {}from $(\phi_0(x), \pi_0(x))$.

In the last section we showed that $U[T]$ exists; here we
explicitly define this operator by giving its action on the
Fock basis of {\S2b}. To begin, we express the Fock ground
state
(Heisenberg vacuum state) as
\begin{equation}
|0,\psi>=\psi({\bf p})\otimes|(+);0>\otimes |(-);0>,
\end{equation}
where $\psi\in{\cal L}^2(R)$. The \Schrodinger picture
image of this state is denoted by 
$|0,\psi;T>$:
\begin{equation}
|0,\psi;T> = U[T] |0,\psi>.
\end{equation}
We note that
\begin{equation}
|0,\psi;T_0> = |0,\psi>.
\end{equation}
To evaluate $ |0,\psi;T>$ it is convenient to decompose 
$U$ as
\begin{equation}
U= V_{\ss\rm II}U_{\ss\rm I} ,
\end{equation}
where $V_{\ss\rm II}$ is the unitary operator
\begin{equation}
V_{\ss\rm II}:= U_{\ss\rm I} U_{\ss\rm II} U_{\ss\rm
I}^{-1},
\end{equation}
and
\begin{equation}
U_{\ss\rm I}=U_{\ss\rm I}^{(+)} U_{\ss\rm I}^{(-)}.
\end{equation}
Using (\ref{eq:UII}) and (\ref{eq:UIq}--\ref{eq:UIa}),
\begin{eqnarray}
V_{\rm\scriptscriptstyle II}[T] &=& 
\exp\Big\{-i
\Big[{p^2\over4\pi}
\int_0^{2\pi}T(x) dx
-
\Big({p\over\sqrt{2}}\sum_{n=1}^\infty
\big[a_{(+)n}Z_n^{(+)}\nonumber\\ 
& &\qquad+a_{(+)n}^\dagger Z_n^{(+)*}
-a_{(-)n}Z_n^{(-)}
-a^\dagger_{(-)n}
Z_n^{(-)*}\big]\Big)\Big]\Big\} ,
\label{eq:VII}
\end{eqnarray}
Our strategy is to first evaluate $U_{\ss\rm I}|0,\psi>$
and then
compute
the action
of $V_{\ss\rm II}$ on the resulting state.
The vector $ U_{\ss\rm I} |0,\psi>$ can be computed
{}from
the observation that it is annihilated by
\begin{eqnarray}
d_{(\pm)k}&:=&U_{\ss\rm I} a_{(\pm)k}U_{\ss\rm I}^\dagger\\
&=&
\sum_{n=1}^\infty\left(\alpha_{(\pm)kn}a_{(\pm)n} + 
\beta_{(\pm)kn}a^\dagger_{(\pm)n}\right),
\end{eqnarray}
where 
\begin{eqnarray}
\alpha_{(\pm)kn}&=&
{1\over2\pi}\sqrt{n\over k}\int_0^{2\pi}e^{ikT^\pm(x)}\,
e^{\mp in x}\,dx \label{eq:alpha}\\ 
\beta_{(\pm)kn}&=&
-{1\over2\pi}\sqrt{n\over k}\int_0^{2\pi}e^{ikT^\pm(x)}\,
e^{\pm in x}\,dx. \label{eq:beta}
\end{eqnarray}

Let us note some important properties of these Bogolubov
coefficients
(see \cite{GSegal} for a more rigorous treatment of most of
these
results). First, note that the operators $d_{(\pm)n}$ can
be obtained {}from (\ref{eq:c}) using the inverse
diffeomorphisms 
$(T^{\pm})^{-1}$:
\begin{equation}
d_{(\pm)n}=c_{(\pm)n}[(T^{\pm})^{-1}].
\end{equation}
The coefficients $\alpha_{(\pm)mn}$ and
$\beta_{(\pm)mn}$ satisfy the relations
\begin{eqnarray}
\sum_{k=1}^\infty\left(\alpha_{(\pm)ik}\alpha^*_{(\pm)jk}-
\beta_{(\pm
)ik}
\beta^*_{(\pm)jk}\right)
&=&\delta_{ij},\\
\sum_{k=1}^\infty\left(\alpha_{(\pm)ik}\beta_{(\pm)jk}-
\beta_{(\pm)ik}
\alpha_{(\pm)jk}\right)
&=&0,
\end{eqnarray}
which are equivalent to saying that the transformation
{\bf(I)} of \S2d is symplectic. The coefficients
$\beta_{(\pm)mn}$ are Hilbert-Schmidt
\begin{equation}
\sum_{m,n=1}^\infty |\beta_{(\pm)mn}|^2 < \infty;
\end{equation}
this result is equivalent to (\ref{eq:HS}). 
%
The
infinite arrays
$\alpha_{(\pm) mn}$ admit inverses $\alpha^{-1}_{(\pm)mn}$
which can be written as
\begin{equation}
\alpha^{-1}_{(\pm)mn}=\alpha^*_{(\pm)nm}-\sum_{k=1}^\infty
\gamma_{(\pm)mk}\beta^*_{(\pm)nk},
\end{equation}
where we have defined the Hilbert-Schmidt operators
\begin{equation}
\gamma_{(\pm)mn}=\sum_{k=1}^\infty
\alpha^{-1}_{(\pm)mk}\beta_{(\pm)kn}.
\end{equation}


It is straightforward to verify that, for any
embedding-dependent function of $p$, $N(p,T)$,
\begin{equation}
U_{\ss\rm I} |0,\psi >=N(p,T)
\exp\Big\{
-{1\over2}
\sum_{k,l=1}^\infty\Big(\gamma_{(+)kl}a^\dagger_{(+)k}
a^\dagger_{(+)l}
+
\gamma_{(-)kl}a^\dagger_{(-)k}a^\dagger_{(-)l}\Big)\Big]
\Big\}|0,\psi>
\label{eq:stateI}
\end{equation}
is annihilated by $d_{(\pm)k}$ for all $k$ (see
\cite{Pilch} for some properties of such a state). Since
$U_{\ss\rm I}$ is trivial on
the
zero mode sector, (\ref{eq:UIq}), (\ref{eq:UIp}),
$N(p,T)$ must be independent of $p$. Thus
\begin{equation}
N(p,T) = N(T),
\end{equation}
and $N(T)$ is determined, up to an embedding-dependent
phase factor, by normalization to be 
\begin{equation}
N(T)=e^{i\Lambda(T)}\det(1-\gamma_{(+)}^*\gamma_{(+)})^{1
\over4}
\det(1-\gamma_{(-)}^*\gamma_{(-)})^{1\over4},
\label{eq:N(T)}
\end{equation}
where $\Lambda(T)$ is an arbitrary real-valued function of
the embedding and we have used a matrix notation in which
 $\gamma_{(\pm)}$ denotes
the symmetric matrix
$\gamma_{(\pm)mn}$. 
$N(T)$ is well-defined
thanks
to the fact that $\gamma$ is Hilbert-Schmidt.

It is now straightforward to compute the action of
$V_{\ss\rm II}$
(\ref{eq:VII})
on (\ref{eq:stateI}) to be
\begin{eqnarray}
|0,\psi,T>&=&M(p,T)\exp\Big\{\sum_{k=1}^\infty\Big[
{-ip \over \sqrt{2}}(\xi_{(+)k} a^\dagger_{(+)k}
+\xi_{(-)k} a^\dagger_{(-)k})\nonumber\\
&-&{1\over2}
\sum_{k,l=1}^\infty\Big(\gamma_{(+)kl}a^\dagger_{(+)k}
a^\dagger_{(+)l}
+
\gamma_{(-)kl}a^\dagger_{(-)k}a^\dagger_{(-)l}\big)\Big]
\Big\}|0,\psi>,
\label{eq:schrodvac}
\end{eqnarray}
where
\begin{equation}
M(p,T) =
\exp\left\{
-i\left[{p^2\over4\pi}\int_0^{2\pi}T(x)dx\right]
 \right\}
\exp\left\{ 
 {p^2\over 4}\sum_{k=1}^\infty
 \big[ 
 \xi_{(+)k}Z^{(+)}_k -\xi_{(-)k}Z^{(-)}_k 
 \big] 
 \right\}
N(T) 
\label{eq:M(T)}
\end{equation}
with $N(T)$ defined by (\ref{eq:N(T)}) and 
\begin{equation}
\xi_{(\pm)k}:=\sum_{l=1}^\infty\alpha^{-1}_{(\pm)kl}
\zeta_{(\pm)l}.
\end{equation}
Note that the various sums and products in the expressions
above 
converge because $\gamma$ is Hilbert-Schmidt and $\xi$, $Z$
are
rapidly 
decreasing. 

The vector $|0,\psi,T>$ serves as the vacuum (or
``cyclic'') vector
for the
Fock representation associated with the annihilation and
creation
operators $b_{(\pm)k}$ and 
$b^\dagger_{(\pm)k}$ where
\begin{eqnarray}
b_{(\pm)k}&:=&U a_{(\pm)k}U^\dagger\\
&=&
i \zeta_{(\pm) k} {p\over \sqrt{2}} +
\sum_{n=1}^\infty\left(\alpha_{(\pm)kn}a_{(\pm)n} + 
\beta_{(\pm)kn}a^\dagger_{(\pm)n}\right) .
\label{eq:b}
\end{eqnarray}
This Fock space representation of the algebra of creation
and annihilation operators and zero modes is
unitarily equivalent to the representation on ${\cal H}$ we
used
originally. By repeatedly applying
the creation operators $b^\dagger_{(\pm)k}$ to
$|0,\psi,T>$, and allowing $\psi$ to range over an
orthonormal basis for ${\cal L}^2(R)$, we obtain an
orthonormal basis $\{|e_i(T)>\}$ for the Hilbert space
${\cal
H}$. This
basis is just the \Schrodinger picture unitary
image of the original orthonormal basis of states
used in the Heisenberg picture. {}From the point of view of
the
parametrized field theory description of \cite{kuchar1} and
\S2c, the 
embedding-independent
 Fock states are the ``physical states'' of the Dirac
quantization based upon the Heisenberg variables. The
physical states
of the Dirac quantization in the \Schrodinger picture are
obtained as
the unitary image of the Heisenberg physical states. The
(pure)
physical states
in the \Schrodinger picture are thus obtained by taking
finite-norm superpositions of the basis $\{|e_i(T)>\}$ for
${\cal H}$
that we described above. The Dirac quantization of the
parametrized
field theory of \cite{kuchar1} in the \Schrodinger picture
is thereby
completed.  However, we would still like to see explicitly
how the
physical states satisfy the quantum constraints in the
\Schrodinger
picture.  This is our next topic.

\subsection*{3b. Functional \Schrodinger equation}

The \Schrodinger picture states constructed in the
last subsection are determined by a choice of embedding. 
In this subsection we consider the change 
induced in these states
by a variation of the embedding. In particular, we derive a
functional \Schrodinger equation that describes the
evolution of the state vector {}from one slice to another of
an arbitrary spacelike foliation. This functional
\Schrodinger
equation is the quantum constraint equation arising in the
Dirac
quantization of parametrized field theory in the
\Schrodinger picture.

To begin, we consider
the embedding dependence of 
the \Schrodinger vacuum state given in
(\ref{eq:schrodvac}), (\ref{eq:N(T)}), (\ref{eq:M(T)}). 
We want to consider the
change induced in this state by an infinitesimal change in
the embedding $T^\alpha(x)$. With this 
result in hand, it is straightforward to compute 
the corresponding results for the basis $\{|e_i(T)>\}$. 
Evidently, we need to compute
the functional derivatives of $\xi_{(\pm)k}$, 
$\gamma_{(\pm)mn}$, and $Z^{(\pm)}_k$ with respect to
$T^\alpha(x)$. 
To display the results of the computation it is convenient
to present the Fourier modes of the functional
derivatives. We define
\begin{equation}
\delta_{(\pm)n}=\int_0^{2\pi}
e^{inT^{\pm}(x)}{\delta\over\delta T^{\pm}(x)} 
\,dx. 
\end{equation}
Direct computation yields
\begin{equation}
\delta_{(\pm)n}\gamma_{(\mp)lm}=0,
\label{eq:deltagamma}
\end{equation}
\begin{equation}
\delta_{(\pm)n}\xi_{(\mp)k}=0,
\end{equation}
\begin{equation}
\delta _{(\pm)n}Z_{(\mp)k}=0,
\end{equation}
\begin{equation}
\delta_{(\pm)n}\gamma_{(\pm)lm}=0\qquad {\rm for}\ n\geq 0,
\end{equation}
\begin{equation}
\delta_{(\pm)n}\gamma_{(\pm)lm}=
-i\sum_{j=1}^{|n|-1}\sqrt{j|n+j|}\alpha^{-1}_{(\pm)lj}
\left[\alpha^*_{(\pm)|n+j|m}
-\sum_{q=1}^\infty\beta^*_{(\pm)|n+j|q}\gamma_{(\pm)qm}
\right],\quad
{\rm for}\ n<0
\end{equation}
\begin{equation}
\delta_{(\pm)n}\xi_{(\pm)k}= 0\qquad{\rm for}\ n\geq0,
\end{equation}	
\begin{eqnarray}
\delta_{(\pm)n}\xi_{(\pm)k}&=&i\sqrt{|n|}\alpha^{-1}_{(\pm)
k|n|}
+i\sum_{j=1}^{|n|-1}\sqrt{j|n+j|}
\alpha^{-1}_{(\pm)kj}\Bigg[\zeta^*_{(\pm)|n+j|}\nonumber\\
&+&\sum_{q=1}^\infty
\beta^*_{(\pm)|n+j|q}\xi_{(\pm)q}\Bigg],\quad{\rm for}
\ n<0,
\label{eq:deltaxi}
\end{eqnarray}
\begin{equation}
\delta _{(\pm)n}Z_{(\pm)k}=\pm i\sqrt{n}\,\alpha_{(\pm)nk}.
\qquad{\rm for}\ n > 0,
\end{equation}
\begin{equation}
\delta _{(\pm)n}Z_{(\pm)k}= 0
\qquad{\rm for}\ n = 0,
\end{equation}
\begin{equation}
\delta _{(\pm)n}Z_{(\pm)k}=\mp
i\sqrt{|n|}\,\beta^*_{(\pm)|n|k}
\qquad{\rm for}\ n<0.
\label{eq:deltaz}\end{equation}

It is now a simple matter to apply $\delta_{(\pm)n}$ to the
state $|0,\psi,T>$ as written in (\ref{eq:schrodvac}),
(\ref{eq:N(T)}), (\ref{eq:M(T)}). The result is a sum of
four terms acting on $|0,\psi,T>$:
\begin{equation}
\delta_{(\pm)n}|0,\psi,T> = \{{\cal P}_{(\pm)n} + {\cal
Q}_{(\pm)n} +
{\cal R}_{(\pm)n} 
+ {\cal S}_{(\pm)n}\}|0,\psi,T>,
\label{eq:deltapsi}
\end{equation}
where ${\cal P}_{(\pm)n}$ is a term proportional
to the identity $\cal I$ arising {}from the derivative of
$N(T)$, 
\begin{equation}
{\cal P}_{(\pm)n} = \delta_{(\pm)n} (\log N(T))\, {\cal I};
\end{equation}
${\cal Q}_{(\pm)n}$ is
quadratic in $p$,
\begin{eqnarray}
{\cal Q}_{(\pm)n} &=& {p^2\over 4}
\Bigg\{
-i({1\over 2\pi} \int_{0}^{2\pi} e^{inT^\pm (x)}dx)
 +\sum_{k=1}^\infty\Big[
\delta_{(\pm)n}\xi_{(+)k}Z_{(+)k}
+\xi_{(+)k}\delta_{(\pm)n}Z_{(+)k}\nonumber\\
\qquad\qquad\qquad && -\delta_{(\pm)n}\xi_{(-)k}Z_{(-)k}
-\xi_{(-)k}\delta_{(\pm)n}Z_{(-)k}
\big]\Bigg\};
\end{eqnarray}
${\cal R}_{(\pm)n}$ is bilinear in $p$ and
$a^\dagger$,
\begin{eqnarray}
{\cal R}_{(\pm)n}&=&- {ip\over\sqrt{2}}\sum_{k=1}^\infty
\left(\delta_{(\pm)n}\xi_{(+)k}
a^\dagger_{(+)k}
+\delta_{(\pm)n}\xi_{(-)k}
a^\dagger_{(-)k}\right);
\end{eqnarray}
and ${\cal S}_{(\pm)n}$ is quadratic in $a^\dagger$,
\begin{eqnarray}
{\cal S}_{(\pm)n}&=&-{1\over2}
\sum_{k,l=1}^\infty\Big(\delta_{(\pm)n}\gamma_{(+)kl}
a^\dagger_{(+)k}
a^\dagger_{(+)l}
+
\delta_{(\pm)n}\gamma_{(-)kl}a^\dagger_{(-)k}a^\dagger_{(-)
l}\big).
\label{eq:T}
\end{eqnarray}

The explicit forms of these terms can be obtained by
substituting 
(\ref{eq:deltagamma})--(\ref{eq:deltaz}). In particular, it
follows
immediately that
for $n\geq 0$
\begin{eqnarray}
{\cal Q}_{(\pm)n}&=&-i{p^2\over4}\delta_{n0},\\
{\cal R}_{(\pm)n}&=&0, \\
{\cal S}_{(\pm)n}&=&0 .
\end{eqnarray}

We
now want to compare these results with the action on
$|0,\psi,T>$ of
the
\Schrodinger picture Hamiltonian. We therefore
digress for a moment to define this Hamiltonian.

The classical dynamical evolution equations 
(\ref{eq:evolphi})--(\ref{eq:pi-dot}) are
generated by the Hamiltonian
\begin{eqnarray}
H&=&\int_0^{2\pi} {1\over4}\Bigg\{{\partial T
^+(x,t)\over\partial t}
(T^+_{,x}(x,t))^{-1}[\pii (x)+\phii_{,x}(x)]^2\nonumber\\
& &-{\partial T^-(x,t)\over\partial t}
(T^-_{,x}(x,t))^{-1}[\pii (x)-\phii_{,x}(x)]^2\Bigg\}\, dx
\label{eq:Hamiltonian}
\end{eqnarray}
Quantum mechanically, the Hamiltonian
(\ref{eq:Hamiltonian}) can be
made
well-defined ({\it i.e.}, densely defined, self-adjoint) 
for any choice of $T^\alpha(x,t)$ by
normal-ordering with respect to the creation and
annihilation operators and
$(a^\dagger,a)$. (This feature does not seem to generalize
to
higher-dimensional models, see \S5).
In this way the normal-ordered Hamiltonian, denoted by
$:H:$, generates
the Heisenberg equations of motion,
\begin{eqnarray}
i{\partial\phi(x)\over\partial t}&=&[\phi(x),:H:]\\
i{\partial\pi(x)\over\partial t}&=&[\pi(x),:H:],
\end{eqnarray}
associated
with an arbitrary spacelike foliation $T^\alpha(x,t)$. 
Because the foliation is arbitrary, the Heisenberg
equations shown
above are equivalent to a
 set of functional Heisenberg equations,
\begin{eqnarray}
i{\delta {\phi} (x) \over \delta T^{\pm}(x^{\prime})}
 & = & 
[\phi(x),{\cal H}_\pm(x^\prime)]\\
i{\delta {\pi} (x) \over \delta T^{\pm}(x^{\prime})}
 & = & 
[\pi(x),{\cal H}_\pm(x^\prime)],
\end{eqnarray}
where
\begin{equation}
{\cal H}_\pm(x) = \pm {:({\pi}(x)\pm {\phi}_{,x}(x))^2:
 \over 4T^{\pm}_{,x}(x)}.
\end{equation}

It is important to keep in mind that normal ordering is
essentially a renormalization prescription that discards an
infinity. It is still possible to renormalize by a finite
amount. This possibility corresponds to the freedom to add
multiples of the identity operator to the Hamiltonian
without disturbing the Heisenberg equations of motion. As we
shall see, this finite renormalization is needed to define
dynamical evolution of the state vector along an arbitrary
foliation.

Recalling the time evolution operator $U[T]$, and the usual
correspondence between the \Schrodinger picture and the
Heisenberg picture, it follows that the time evolution of
state vectors is (up to the possible addition of multiples
of the identity) controlled by the \Schrodinger Hamiltonian,
\begin{equation}
H_{\scriptscriptstyle S}:=U[T] :H: U^\dagger[T],
\end{equation}
and \Schrodinger Hamiltonian densities,
\begin{equation}
{\cal H}_{{\scriptscriptstyle S}\pm}(x):=U[T] {\cal
H}_\pm(x)
U^\dagger[T].
\end{equation}
{}From the definition (\ref{eq:b}) of the operators
$b_{(\pm)k}$ and
$b_{(\pm)k}^\dagger$, it is straightforward to verify that
$H_{\scriptscriptstyle S}$ and ${\cal
H}_{{\scriptscriptstyle S}\pm}$ are the same functions of
$b_{(\pm)k}$ and $b_{(\pm)k}^\dagger$ that $:H:$ and ${\cal
H}_\pm$ are functions of $a_{(\pm)k}$ and
$a_{(\pm)k}^\dagger$. In
particular, the \Schrodinger Hamiltonians
$H_{\scriptscriptstyle S}$
and ${\cal
H}_{{\scriptscriptstyle S}\pm}$ are 
normal-ordered in the $b$, $b^\dagger$ operators.

We now return to our derivation of the functional
\Schrodinger equation satisfied by $|0,\psi,T>$. To this
end, we consider the action of the operators ${\cal
H}_{{\scriptscriptstyle S}\pm}(x)$ on $|0,\psi,T>$. Again,
we
introduce Fourier modes:
\begin{equation}
h_{(\pm)n}=\int_0^{2\pi} e^{inT^{\pm}(x)}
{\cal H}_{{\scriptscriptstyle S}\pm}(x)\,dx.
\end{equation}
These Fourier modes are Virasoro
operators (familiar {}from string theory) built {}from the
$b$,
$b^\dagger$ operators: 
\begin{equation}
h_{(\pm)0}={p^2\over 4}+\sum_{k=1}^\infty
k\left(b^\dagger_{(\pm)k}
b_{(\pm)k}\right),
\end{equation}
and, for $n>0$,
\begin{eqnarray}
h_{(\pm)n}&=&
-i\sqrt{{n\over2}}p\,b_{(\pm)n}
+\sum_{k=1}^\infty\sqrt{k(k+n)}b_{(\pm)k}^\dagger
b_{(\pm)k+n}\nonumber\\
&-&{1\over2}\sum_{k=1}^{n-1}\sqrt{k(n-k)}b_{(\pm)k}b_{(\pm)
n-k},\\
\bigskip
h_{(\pm)-n} &=& h^\dagger_{(\pm)n}.
\end{eqnarray}

We now compute the action of $h_{(\pm)n}$ on $|0,\psi,T>$
in order
to compare with (\ref{eq:deltapsi}). To begin we note that,
because
this state is the vacuum associated with the $(b_{(\pm)n},
b_{(\pm)n}^\dagger)$ operators, we have
\begin{equation}
h_{(\pm)n} |0,\psi,T> = \delta_{n0}{p^2\over4}
|0,\psi,T>\quad n\geq0.
\end{equation}

Using (\ref{eq:deltagamma})--(\ref{eq:deltaxi}), 
(\ref{eq:deltapsi})--(\ref{eq:T}) we see that
\begin{equation}
\left[{1\over i}\delta_{(\pm)n} + h_{(\pm)n} 
+ i(\delta_{(\pm)n} \log N(T)) {\cal I}\right]|0,\psi,T> =
0,\quad
n\geq0.
\end{equation}
Thus, up to addition of a multiple of the identity to the
\Schrodinger
Hamiltonian, we have obtained the expected functional
\Schrodinger
equation for $n\geq0$.

In order to compute the action of
$h_{(\pm)-n}=h_{(\pm)n}^\dagger$ on 
$|0,\psi,T>$ we expand the $(b_{(\pm)n},
b_{(\pm)n}^\dagger)$
operators in
terms of the 
$(a_{(\pm)n}, a_{(\pm)n}^\dagger)$ operators using the
Bogolubov
transformation 
(\ref{eq:b}) and apply the resulting operator to
$|0,\psi,T>$. At
this point it is convenient to take note of the identities
\begin{equation}
\alpha^{-1}_{(\pm)kl}=
\alpha^*_{(\pm)lk}
-\sum_{r=1}^\infty \beta^*_{(\pm)lr}\gamma_{(\pm)rk},
\end{equation}
\begin{equation}
\sum_{k=1}^\infty \alpha^{-1}_{(\pm)kl} Z_{k}
=\mp
(\zeta^*_{(\pm)l}+\sum_{k=1}^\infty\beta^*_{(\pm)lk}
\xi_{(\pm)k}).
\end{equation}

We get four types of terms:
\begin{equation}
h^\dagger_{(\pm)n}|0,\psi,T> = \left(P_{(\pm)n} +
Q_{(\pm)n} +
R_{(\pm)n}
+S_{(\pm)n}\right)|0,\psi,T>.
\end{equation}
Here $P_{(\pm)n}$ is proportional to the identity $\cal I$,
\begin{equation}
P_{(\pm)n} = - {1\over2} \sum_{j=1}^{n-1}\sum_{r=1}^\infty
\sqrt{j(n-j)}
\beta^*_{(\pm)jr}\alpha^{-1}_{(\pm)r,n-j}\,{\cal I},
\end{equation}
$Q_{(\pm)n}$ is quadratic in $p$,
\begin{eqnarray}
Q_{(\pm)n}&=&{p^2\over2}\Bigg\{
\sqrt{n}\zeta^*_{(\pm)n}
+\sqrt{n}\sum_{l=1}^\infty\beta_{(\pm)nl}^* \xi_{(\pm)l}
+\sum_{k=1}^{n-1}\sqrt{k(n-k)}\Big[{1\over2}\zeta^*_{(\pm)k}
\sum_{l=1}^\infty\beta^*_{(\pm)n-k,l}\xi_{(\pm)l}\nonumber\\
&+&
{1\over2}\zeta^*_{(\pm)n-k}
\sum_{l=1}^\infty\beta^*_{(\pm)k,l}\xi_{(\pm)l}+
{1\over2}\zeta^*_{(\pm)k}
\zeta^*_{(\pm)n-k} + {1\over2}\sum_{l,m=1}^\infty
\beta^*_{(\pm)kl}\beta^*_{(\pm)n-k,m}\xi_{(\pm)l}
\xi_{(\pm)m}\Big]\Bigg\}.\nonumber\\
\end{eqnarray}
$R_{(\pm)n}$ is bilinear in $p$ and $a_{(\pm)n}^\dagger$:
\begin{eqnarray}
R_{(\pm)n}&=&
i \sqrt{{n\over2}}p \sum_{j=1}^\infty (\alpha^*_{(\pm)nj}
-\sum_{r=1}^\infty
\beta^*_{(\pm)nr}\gamma_{(\pm)rj})a^\dagger_{(\pm)j}
\nonumber\\
&+&{i\over\sqrt{2}} p
\sum_{j=1}^\infty\sum_{k=1}^{n-1}\sqrt{k(n-k)}
 a^\dagger_{(\pm)j}\alpha^{-1}_{(\pm)jk}\Bigg\{
\zeta^*_{(\pm)n-k}+\sum_{l=1}^\infty\beta^*_{(\pm)n-k,l}\xi
_{(\pm)l}
\Bigg\}.
\end{eqnarray}
Finally, $S_{(\pm)n}$ is quadratic in $a^\dagger_{(\pm)k}$:
\begin{eqnarray}
& & S_{(\pm)n}
=
-{1\over2}\sum_{l,m=1}^\infty\sum_{k=1}^{n-1}\sqrt{k(n-k)}
\Bigg[\alpha^*_{(\pm)kl}\alpha^*_{(\pm)n-k,m}
-\sum_{r=1}^\infty\beta^*_{(\pm)kr}\alpha^*_{(\pm)n-k,l}
\gamma_{(\pm)
rm}
\nonumber\\
&\quad-&\sum_{r=1}^\infty\beta^*_{(\pm)n-k,r}\alpha^*_{(\pm
)k,l}
\gamma_{(\pm)rm}
+\sum_{r,s=1}^\infty\beta^*_{(\pm)kr}\beta^*_{(\pm)n-k,s}
\gamma_{(\pm
)rl}
\gamma_{(\pm)sm}\Bigg]a^\dagger_{(\pm)l}a^\dagger_{(\pm)m}
\end{eqnarray}

We now compare $Q_{(\pm)n}$, $R_{(\pm)n}$, $S_{(\pm)n}$
with ${\cal
Q}_{(\pm)n}$, 
${\cal R}_{(\pm)n}$, ${\cal S}_{(\pm)n}$; we find that
\begin{eqnarray}
Q_{(\pm)n}&=&i{\cal Q}_{(\pm)-|n|}\\
R_{(\pm)n}&=&i{\cal R}_{(\pm)-|n|}\\
S_{(\pm)n}&=&i{\cal S}_{(\pm)-|n|}.
\end{eqnarray}

Combining our results, we have for all $n$
\begin{equation}
\left[{1\over i}\delta_{(\pm)n} + h_{(\pm)n} 
+ {\cal A}_{(\pm)n}{\cal I}\right]|0,\psi,T> = 0,
\end{equation}
where
\begin{eqnarray}
{\cal A}_{(\pm)n}
&=&
 i(\delta_{(\pm)n} \log N(T)),
\phantom{+ {1\over2} \sum_{j=1}^{n-1}\sum_{r=1}^\infty}
\qquad {\rm when}\ n\geq 0,\\
&=&i(\delta_{(\pm)n} \log N(T))\nonumber\\
&+& {1\over2}
\sum_{j=1}^{n-1}\sum_{r=1}^\infty \sqrt{j|n+j|}
\beta^*_{(\pm)jr}\alpha^{-1}_{(\pm)r,|n|-j},\quad
{\rm when}\
n<0.
\end{eqnarray} 
This equation is equivalent to
\begin{equation}
\left[{1\over i}{\delta\over\delta T^\alpha(x)}
+{\cal H}_{\scriptscriptstyle S\alpha}(x) + {\cal
A}_\alpha(x){\cal I}
\right]|0,\psi,T>=0,
\label{eq:schrod}
\end{equation}
where
\begin{equation}
{\cal A}_\pm(x) =
{1\over2\pi}T^\pm_{,x}(x)\sum_{n=-\infty}^\infty
e^{-inT^\pm(x)}
{\cal A}_{(\pm)n}.
\end{equation}

The presence of the $c$-number contribution ${\cal
A}_\alpha$ to the
\Schrodinger picture image of the normal-ordered Heisenberg
Hamiltonian was proposed by Kucha\v r in \cite{kuchar1}. 
Its presence is needed
to ensure the integrability of 
(\ref{eq:schrod}) given the appearance of an anomaly
(Schwinger
terms) in the algebra of the operators ${\cal H}_\alpha(x)$.
As such,
following Kucha\v r, we refer to ${\cal A}_\alpha$ as the
``anomaly
potential''.  The form of ${\cal A}_\alpha$ as a functional
of
embeddings is not uniquely determined because of the
freedom to
specify 
$\Lambda[T]$ in (\ref{eq:N(T)}).   The results of
\cite{kuchar1} imply that the
phase $\Lambda[T]$ can be chosen to put the anomaly
potential into
the following local, spatially covariant form
\cite{sign_error}: 
\begin{equation}
{\cal A}_\pm ={1\over
24\pi}\left[\mp{1\over2}(T^\pm_{,x})^{-1}
+\left((T^\pm_{,x})^{-1}K_x\right)_{,x}\right],
\label{eq:acovariant}
\end{equation}
where
\begin{equation}
K_x={1\over2}\left[{T^-_{,xx}\over T^-_{,x}}
-{T^+_{,xx}\over T^+_{,x}}\right]
\label{eq:extrinsic}
\end{equation}
is the mean extrinsic curvature of the embedding multiplied
by the
square root of the determinant of the metric induced on the
embedded
circle.

Having derived the functional \Schrodinger equation
satisfied by the
\Schrodinger image of the Heisenberg vacuum state, it now is
easy to
see that the basis $\{|e_i(T)>\}$
described in
\S3a also satisfies the same equation.  This follows {}from
the fact that
the operators $p$, $b_{(\pm)k}$, 
$b^\dagger_{(\pm)k}$, $k=1,2,...$ satisfy
\begin{equation}
[p,{1\over i}{\delta\over\delta T^\alpha(x)}
+{\cal H}_{\scriptscriptstyle S\alpha}(x) + {\cal
A}_\alpha(x){\cal
I}]
=0,
\label{eq:observablep}
\end{equation}
and
\begin{equation}
[b_{(\pm) k},{1\over i}{\delta\over\delta T^\alpha(x)}
+{\cal H}_{\scriptscriptstyle S\alpha}(x) + {\cal
A}_\alpha(x){\cal
I}]
=0=
[b^\dagger_{(\pm) k},{1\over i}{\delta\over\delta
T^\alpha(x)}
+{\cal H}_{\scriptscriptstyle S\alpha}(x) + {\cal
A}_\alpha(x){\cal
I}].
\label{eq:observableb}
\end{equation}
The states $ \{|e_i(T)>\}$ thus define a basis of solutions
to
the
functional \Schrodinger equation.

Finally, we emphasize that the functional \Schrodinger
equation
(\ref{eq:schrod}) can be viewed as the quantum constraint
in the
Dirac quantization of parametrized field theory in the
\Schrodinger
picture.  As predicted in \cite{kuchar1}, the factor
ordering of this constraint
is quite non-trivially related to that of normal ordering
in the
$(a^\dagger,a)$ operators.  Note also that the operators
$(p,b^\dagger,b)$ used to build the physical states are
``Dirac
observables''; 
as shown in (\ref{eq:observablep}) and 
(\ref{eq:observableb}) 
they commute with quantum constraint operators.

\subsection*{4. Spatial Diffeomorphisms}

In the quantum theory of generally covariant systems one
often
partitions the constraint equations of the theory into
dynamical
constraints (the ``super-Hamiltonian constraint'', the
``Wheeler-DeWitt equation'') and gauge constraints (the
``super-momentum constraint'', the ``diffeomorphism
constraint'').  
The physical states constructed in \S3a satisfy the
functional
\Schrodinger equation (\ref{eq:schrod}), which governs the
propagation
of the state vector {}from hypersurface to hypersurface in
spacetime. As
described in \S2c, this equation can be interpreted as
representing a
quantization of the constraints which arise in the
Hamiltonian
description of a parametrized field theory.  If equation
(\ref{eq:schrod}) is projected along the normal to the
embedding
$T^\alpha(x)$ then we obtain an analog of the Wheeler-DeWitt
equation, which governs the change of the state as time is
pushed
forward along the normal to the embedding. If we project
this
equation tangentially to the embedding $T^\alpha(x)$, then
we get
\begin{equation}
\left[{1\over i}T^\alpha_{,x}{\delta\over\delta
T^\alpha} + {\cal
H}_{{\ss (S)}x}
+ {\cal A}_x\right]|\Psi(T)>=0,
\label{eq:diffeo}
\end{equation}
where
\begin{equation}
{\cal H}_{{\ss (S)}x}=T^\alpha_{,x} {\cal H}_{{\ss
(S)}\alpha},
\end{equation}
and
\begin{equation}
{\cal A}_{x}=T^\alpha_{,x} {\cal A}_{\alpha}.
\end{equation}
Normally, this gauge constraint is viewed as enforcing some
kind of
spatial diffeomorphism invariance of the state vector. 
Indeed, the
analog of this equation in canonical quantum gravity is
usually
interpreted as saying that wavefunctions in the metric
representation
depend only upon diffeomorphism
equivalence classes
of the spatial metric \cite{DeWitt}.  Alternatively, in the
loop representation
of canonical quantum gravity, the analog of 
(\ref{eq:diffeo}) is interpreted as saying that
wavefunctions only
depend upon diffeomorphism equivalence classes of closed
curves
(knots, links, etc.) \cite{Rovelli,Ashtekar}.  Here we
would like to
relate
(\ref{eq:diffeo}) to the action of spatial diffeomorphisms
in quantum
parametrized field theories.  In particular, we would like
to see
how/if one can maintain the interpretation of
(\ref{eq:diffeo}) as
enforcing spatial diffeomorphism invariance at the quantum
level. 
The issue is not trivial given the factor ordering
used to
define ${\cal H}_{{\ss (S)}x}$ and, in particular, given
the $c$-number
term ${\cal A}_{x}$ which appears in (\ref{eq:diffeo}).

We will present two results.  First we show that the phase
freedom
($\Lambda[T]$ in (\ref{eq:N(T)})) can be used to cast
(\ref{eq:diffeo}) into the form
\begin{equation}
\left[{1\over i}T^\alpha_{,x}{\delta\over\delta
T^\alpha} +
h_x\right]
|\Psi(T)>=0,
\label{eq:newdiffeo}
\end{equation}
where
\begin{equation}
h_x=:\pi_0 (\phi_0),_x:,
\label{eq:h1}
\end{equation}
is a particular ordering of the \Schrodinger picture
momentum density
for the field, and the field operators 
$\phi_0(x)$ and $\pi_0(x)$ are defined in (\ref{eq:phi0}),
(\ref{eq:pi0}).  By definition, the operator $h_x$ is
normal ordered
in the $(a^\dagger,a)$ creation and annihilation
operators.  Second,
we show equation (\ref{eq:newdiffeo}) can be interpreted as
indicating that the physical states constructed in \S3a are
invariant
under an action of the group of (spatial) diffeomorphisms
of the
circle.

To begin, we note that ${\cal H}_{{\ss (S)}x}$ is, up to
operator
ordering, the \Schrodinger momentum density in
(\ref{eq:h1}).  As a
consequence, the difference between ${\cal H}_{{\ss
(S)}x}(x)$
and $h_x(x)$
is a ``$c$-number'' functional of the embeddings, 
$\sigma[T](x)$:
\begin{equation}
{\cal H}_{{\ss (S)}x}=h_x + \sigma{\cal I}.
\label{eq:defsigma}
\end{equation}
A direct computation of this $c$-number is straightforward
but not
immediately enlightening.  We compute $\sigma[T](x)$
indirectly as
follows.  Because of 
(\ref{eq:defsigma}), the variation of 
${\cal H}_{{\ss (S)}x}$ with respect to the embedding
$T^\alpha(x)$
is a multiple of the identity which is related to
$\sigma[T]$ via
\begin{equation}
{\delta {\cal H}_{{\ss (S)}x}[T](x)\over\delta T^\alpha(y)}
={\delta \sigma[T](x)\over\delta T^\alpha(y)} {\cal I}.
\end{equation}
 We take the expectation value of this operator relation in
the
\Schrodinger vacuum state $|0,\psi,T>$.  Using the
\Schrodinger
equation (\ref{eq:schrod}) we can put the expectation value
in the
form
\begin{eqnarray}
{\delta \sigma[T](x)\over\delta T^\alpha(y)}&=&
iT^\beta_{,x}(x)
<0,\psi|\,[{\cal H}_{\beta}(x),
{\cal H}_{\alpha}(y)]|0,\psi>\nonumber\\
&-&i{\delta\over\delta
T^\alpha(y)}<0,\psi|T^\beta_{,x}(x){\cal
H}_\beta(x)
\,|0,\psi>.
\label{eq:dsigma1}
\end{eqnarray}

The right-hand side of (\ref{eq:dsigma1}) can be evaluated
using
results of Kucha\v r \cite{kuchar1}.   As usual, we will
compute in null
coordinates; we have \cite{sign_error}
\begin{equation}
{\delta \sigma[T](x)\over\delta T^\pm(y)}
=\pm{1\over24\pi}T^\pm_{,x}(x)
\left\{\delta_{,x}(x,y)
+\partial_x\left[(T^\pm_{,x}(x))^{-1}\partial_x\left((T^\pm_
{,x}(x))^{
-1}
\delta_{,x}(x,y)\right)\right]\right\}.
\label{eq:dsigma2}
\end{equation}
  It is a straightforward exercise
to solve the
functional differential equation (\ref{eq:dsigma2}); we get
\begin{eqnarray}
\sigma[T]&=&{1\over24\pi}
\Bigg[{1\over2}(T^+_{,x})^2 - {3\over2}(T^+_{,x})^{-2}
(T^+_{,xx})^2
+(T^+_{,x})^{-1}T^+_{,xxx}\nonumber\\
&&-{1\over2}(T^-_{,x})^2 + {3\over2}(T^-_{,x})^{-2}
(T^-_{,xx})^2
-(T^-_{,x})^{-1}T^-_{,xxx}
\Bigg],
\label{eq:schwarz}
\end{eqnarray}
where we have eliminated an integration constant by taking
into
account the boundary condition that $\sigma[T]=0$ when
$T^\alpha(x)=T^\alpha_0(x)$.

As mentioned in \S2d, the dynamical evolution of field
operators
arises via two copies of the metaplectic representation of
the group
of diffeomorphisms of the circle. As noted in
\cite{GSegal}, this
representation is closely related to a version of the
Schwarzian
derivative.   The Schwarzian derivative defined in
\cite{GSegal} is a
non-linear third-order differential operator mapping
diffeomorphisms
of the circle into functions on the circle.  It is defined
on
diffeomorphisms $f\colon S^1\to S^1$ via
\begin{equation}
S(f)={1\over12}{f^{\prime\prime\prime}\over
f^\prime}-{1\over8}
\left({f^{\prime\prime}\over f^\prime}\right)^2
+{1\over24}[(f^\prime)^2-1].
\end{equation}
The difference between the two different orderings of the
\Schrodinger
momentum densities can therefore be expressed in terms of
the Schwarzian derivative as
\begin{equation}
\sigma[T]={1\over2\pi}\left[S(T^+)-S(T^-)\right].
\end{equation}

{}From the result (\ref{eq:schwarz}), it is now easy to
show that, for
an appropriate choice of $\Lambda[T]$ in (\ref{eq:N(T)}),
we can turn
(\ref{eq:diffeo}) into (\ref{eq:newdiffeo}), {\it i.e.},
\begin{equation}
{\cal A}_x[T] + \sigma[T] = 0.
\label{eq:acancel}
\end{equation}
Indeed, the local, spatially covariant choice of ``gauge'' 
advocated by Kucha\v r in \cite{kuchar1} leads precisely to
(\ref{eq:acancel}).  This is easily verified using
(\ref{eq:acovariant}), and then using the relation between
the
extrinsic curvature and the embeddings
(\ref{eq:extrinsic}).  We thus
get an interpretation of Kucha\v r's covariant choice of
gauge:  In
this gauge the anomaly potential exactly compensates for the
difference in factor ordering between the \Schrodinger
momentum
density ${\cal H}_{{\ss (S)}x}$ appearing in
(\ref{eq:schrod}) and
the naive \Schrodinger momentum density (\ref{eq:h1}).

Given an appropriate choice of phase $\Lambda[T]$ in
(\ref{eq:N(T)}),
we can assume that the spatial projection of the functional
\Schrodinger equation takes the form (\ref{eq:newdiffeo}). 
We now
show that this equation implies spatial diffeomorphism
invariance of
the \Schrodinger picture physical states.  Although this
could be
demonstrated directly in the Fock representation we have
been using
for the non-zero modes of the field, we will instead place
our
discussion in the \Schrodinger coordinate representation
since that is
the representation one usually has in mind in such
discussions.  We
now digress to describe this representation.

The \Schrodinger representation we shall use is a natural
extension to
infinitely many degrees of freedom of an analogous
representation for
the harmonic oscillator.  Because of the
absence of an 
infinite-dimensional generalization of the usual
translationally
invariant Lebesgue measure,  we use a Gaussian measure
$d\mu$ to
define the Hilbert space inner product \cite{Glimm, Segal}. 
So, the Hilbert space $\cal H$ of states is defined as a
space of
functionals 
$\Psi=\Psi[Q]$ of a scalar field $Q(x)$ on a circle.  We
assume that
the scalar field lies in the function space which is the
topological dual 
to the space of smooth functions on the circle.
Thus $Q(x)\in
{\cal S}^\prime$, the space of  distributions on the
circle (see {\it e.g.}, \cite{Choquet}). 
It is convenient to work with the Fourier modes
of $Q(x)$.  We have 
\begin{equation}
Q(x)=\sum_{n=-\infty}^\infty Q_n e^{-inx},
\end{equation}
and, since $Q(x)$ is real,
\begin{equation}
Q_n = Q_{-n}^* .
\end{equation}

The scalar product $(\cdot,\cdot)$ on $\cal H$ is that
associated with the Gaussian
measure
$d\mu[Q]$ on the space of fields $Q(x)$ with covariance
${1\over\pi}
\left(-{d^2\over dx^2}\right)^{-1/2}$ for the non-zero
modes of
$Q(x)$.  The zero mode $Q_0$ gets the standard
translationally
invariant measure $dQ_0$.  So, for example, if we consider
wavefunctions depending upon a finite number of modes, say, 
$\{Q_n, |n|\leq N\}$, we have
\begin{equation}
(\Psi,\Phi)=\int \Psi^*[Q] \Phi[Q] dQ_0 
\prod_{n=-N}^{\ \ N\ \star}{|2n|^{1/2}\over\pi^{1/2}}
e^{-|n|Q_nQ_{-n}} 
dQ_n .
\label{eq:innerproduct}
\end{equation}
Here the star on the product symbol indicates one should
omit
$n=0$.  The Hilbert space inner product based upon the
Gaussian
measure $d\mu[Q]$ arises formally as the limit of
(\ref{eq:innerproduct}) as 
$N\rightarrow\infty$.
   
Because we use the measure $d\mu[Q]$, the wave functions
$\Psi[Q]$
cannot be quite interpreted as probability
amplitudes in the
traditional way.  Note, for example, that the Fock vacuum
$|0,\psi>$
in this representation is simply given by the wavefunction
$\Psi[Q]=\psi(Q_0)$, where $\psi\in {\cal L}^2(R)$.  
In general, if the wavefunction is given by $\Psi=\Psi[Q]$,
the
probability  ${\cal P}[Q]$ for measuring the field 
$\phi(x)$ and obtaining a value (in an infinitesimal
neighborhood of) 
$Q(x)$ is given by
\begin{equation}
{\cal P}[Q]=\Psi^*[Q]\Psi[Q]d\mu[Q].
\label{eq:density1}
\end{equation}
Inclusion of the Gaussian measure in (\ref{eq:density1}) is
essential
for the probability interpretation of the wavefunctions.  

Keeping in mind that the Heisenberg picture fields on the
initial
slice $X^\alpha_0(x)$, namely $(\phi_0(x),\pi_0(x))$, are
the
\Schrodinger picture fields, we expand these operators as
\begin{equation}
\phi_0(x)={1\over \sqrt{2\pi}}\sum_{n=-\infty}^\infty
\phi_n e^{-inx},
\end{equation}
\begin{equation}
\pi_0(x)={1\over \sqrt{2\pi}}\sum_{n=-\infty}^\infty \pi_n
e^{inx}.
\end{equation}
The Fourier representatives $(\phi_n,\pi_n)$ of the
\Schrodinger
picture operators $(\phi_0(x),\pi_0(x))$ are to satisfy the
commutation relations
\begin{equation}
[\phi_n,\pi_m] = i \delta_{n,m},
\end{equation}
and the Hemiticity requirements
\begin{equation}
\phi^\dagger_n=\phi_{-n}\quad {\rm and}\quad
\pi^\dagger_n=\pi_{-n}.
\end{equation}
The basic operators $(\phi_n,\pi_n)$ are represented
on wavefunctions as
\begin{equation}
\phi_n\Psi[Q] = Q_n \Psi[Q],
\end{equation}
\begin{equation}
\pi_n\Psi[Q]={1\over i}\left({\partial\Psi[Q]\over\partial
Q_{n}}
-|n|Q_{-n}
\Psi[Q]
\right).
\end{equation}
The creation and annihilation operators are represented as
\begin{eqnarray}
a_{(\pm)n}\Psi[Q]&=&{1\over\sqrt{ 2n}}{\partial\Psi\over 
\partial Q_{\mp n}}\\
a^\dagger_{(\pm_)n}\Psi[Q]&=&
-{1\over\sqrt{2 n}}{\partial\Psi\over 
\partial Q_{\pm n}} +  \sqrt{2n}Q_{\mp n}
\Psi.
\end{eqnarray}
The \Schrodinger representation described here is unitarily
equivalent
to the Fock representation \cite{Glimm, Segal}.

It is now a simple matter to express the \Schrodinger
momentum density 
(\ref{eq:h1}) as a differential operator-valued
distribution on a
suitable dense domain of functions $\Psi[Q]$.  We get
\begin{eqnarray}
h_x(x) \Psi[Q]&=& -{1\over2\pi}\sum_{n,m=-\infty}^\infty 
e^{i(n-m)x}m Q_m{\partial\Psi[Q]\over\partial Q_n}
\nonumber\\
&-&{1\over 2\pi}\sum_{n=1}^\infty 
 \sum_{m=1}^{n}  m (n-m)
\left[ e^{inx}Q_{-(n-m)}Q_{-m} - 
e^{-inx}Q_{n-m}Q_{m}\right]\Psi[Q].\nonumber\\
&&
\label{eq:haction}
\end{eqnarray}

We now consider an action of the spatial diffeomorphism
group
Diff$(S^1)$ on state vectors in this representation.  Let
$f\colon
S^1\to S^1$ be a diffeomorphism of the circle.  In
coordinates, $f$
is represented by a smooth map $x\longrightarrow f(x)$ with
a smooth
inverse, satisfying
\begin{equation}
f(2\pi)=f(0)+2\pi.
\end{equation}
We consider the usual pull-back action of spatial
diffeomorphisms on
the field $Q(x)$:
\begin{equation}
Q(x)\longrightarrow (f^*Q)(x) := Q(f(x)).
\end{equation}
This action induces an action of Diff$(S^1)$ on the Fourier
modes:
\begin{equation}
Q_n\longrightarrow (f^*Q)_n := \sum_{m=-\infty}^\infty
\Xi_{nm}Q_m,
\end{equation}
where
\begin{equation}
\Xi_{nm}={1\over2\pi} \int_0^{2\pi} e^{inx}e^{-imf(x)}\, dx.
\end{equation}
In order to interpret (\ref{eq:newdiffeo}) we need the
infinitesimal
form of this action.  Consider a 1-parameter family
$f_\lambda$ of
spatial diffeomorphisms and define
\begin{equation}
V(x):=\left({df_\lambda(x)\over d\lambda}\right)_{\lambda=0}
=\sum_{n=-\infty}^\infty V_ne^{-inx},
\end{equation}
\begin{equation}
\delta Q_n:= \left({d(f_\lambda^*Q)_n\over
d\lambda}\right)_{\lambda=0}.
\end{equation}
It is easy to see that
\begin{equation}
\delta Q_n=- i\sum_{m=-\infty}^\infty m\,V_{n-m} Q_m.
\end{equation}

Let us now define an operator $\delta_{V}$ which provides
the
infinitesimal action of a one parameter family of spatial
diffeomorphisms $f_\lambda$ generated by $V$ on functionals
$\Psi[Q]$:
\begin{equation}
\delta_V \Psi[Q] = 
\left({d \Psi[f^*_\lambda Q]\over
d\lambda}\right)_{\lambda=0}.
\end{equation}
If we also define
\begin{equation}
h_x(V) = \int_0^{2\pi} h_x(x) V(x)\,dx,
\end{equation}
then, using (\ref{eq:haction}), it is easily verified that
\begin{equation}
h_x(V)\Psi[Q]={1\over i}\delta_V\Psi[Q] + F[Q]\Psi[Q],
\label{eq:h1psi}
\end{equation}
where
\begin{equation}
F[Q] = -\sum_{n=1}^\infty 
 \sum_{m=1}^{n} m (n-m)
\left[ V_n Q_{-(n-m)}Q_{-m} - 
V_{-n}Q_{n-m}Q_{m}\right] .
\end{equation}
We remark that the infinite sum in $F[Q]$ converges 
for sufficiently smooth $V(x)$.  
{}From (\ref{eq:h1psi}) we see that $h_x(V)$ 
would generate the action of spatial
diffeomorphisms on wavefunctions $\Psi[Q]$ 
if not for the
presence of
the term $F[Q]$.  This extra term simply reflects the
presence of the
Gaussian measure in (\ref{eq:innerproduct}).  The role of
$F[Q]$ is
to guarantee that  $h_x(V)$ generates the action of spatial
diffeomorphisms on the probability  
(\ref{eq:density1}).  Indeed, we have the identity
\begin{equation}
\delta_V{\cal P}=\Big\{[ih_x(V)\Psi]^*\Psi +
\Psi^*[ih_x(V)\Psi]
\Big\}d\mu.
\label{eq:dp}
\end{equation}

Next we recall that a functional $\Phi[T]$ of the
embeddings changes
under an infinitesimal spatial diffeomorphism via
\begin{equation}
\left({d \Phi[T^\alpha\circ f_\lambda]\over
d\lambda}\right)_{\lambda=0}
=\int_0^{2\pi} {\delta \Phi\over\delta T^\alpha(x)}
T^\alpha_{,x}(x)
V(x) \,dx.
\label{eq:deltat}
\end{equation}
Because of (\ref{eq:dp}), the spatial projection of the
functional
\Schrodinger equation, given in (\ref{eq:newdiffeo}), then
implies
that the probabilities occurring on a given embedding are
invariant
under orientation preserving spatial diffeomorphisms.  More
precisely, associated with a physical state vector, such as
(\ref{eq:schrodvac}), there is a wavefunction
\begin{equation}
\Psi = \Psi[Q,T]
\end{equation}
which defines the probability ${\cal P}[Q,T]$ for a
measurement of the field $\phi(x)$ on the circle embedded as
$T^\alpha=T^\alpha(x)$ to result in $Q(x)$:
\begin{equation}
{\cal P}[Q,T] = \Psi^*[Q,T] \Psi[Q,T] d\mu[Q].
\end{equation}
The probability ${\cal P}[Q,T]$ is spatially diffeomorphism
invariant: 
If $f\colon S^1\to S^1$ is an orientation-preserving
diffeomorphism, then
\begin{equation}
{\cal P}[Q,T] = {\cal P}[f^*Q,T\circ f].
\label{eq:invariance}
\end{equation}
The result (\ref{eq:invariance}) is checked as follows.
Because any two orientation preserving diffeomorphisms of
the circle
can be connected by a one parameter family of such
diffeomorphisms,
it suffices to consider a one parameter family of
diffeomorphisms in
(\ref{eq:invariance}) and check that
\begin{equation}
\left({d{\cal P}[f_\lambda^*Q,T\circ f_\lambda]\over
d\lambda}
\right)_{\lambda=0}=0.
\label{eq:deltap}
\end{equation}
Using (\ref{eq:deltat}), (\ref{eq:dp}), and
(\ref{eq:newdiffeo}),
equation 
(\ref{eq:deltap}) follows.

We note that while equation (\ref{eq:newdiffeo}) depends
upon the
choice of phase 
$\Lambda[T]$, the result (\ref{eq:invariance}) is
independent of such
a choice of phase.  This is, of course, due to the fact
that the phase
factor does not contribute to the probability. 
Viewing the
state of a quantum system as the totality of probability
distributions for the outcome of any and all measurements
made on an
ensemble of identically prepared systems, we thus conclude
that the
functional \Schrodinger equation (\ref{eq:schrod}) enforces
spatial
diffeomorphism invariance of states in the \Schrodinger
representation
of the \Schrodinger picture.

Physically speaking, there is little else to discuss
regarding the
role of spatial diffeomorphisms in the space of \Schrodinger
picture
physical states.  Mathematically, there are a few other
interesting
issues.  In particular, while the probabilities are
spatially
diffeomorphism invariant in the sense of
(\ref{eq:invariance}), in the
present representation neither the measure 
$d\mu[Q]$ nor the wavefunctions 
$\Psi[Q,T]$ satisfying (\ref{eq:schrod}) are separately
invariant
under the spatial diffeomorphism transformation
\begin{equation}
(Q,T)\longrightarrow (f^*Q,T\circ f).
\end{equation}
This is because the representation we are working in is
designed to
render the initial field operators (the \Schrodinger picture
field
operators) diagonal and keep in a simple form the
representation of
the $(a^\dagger,a)$ creation and annihilation operators as
well as
the representation of the Fock vacuum 
$|0,\psi>$.  {}From the point of view of the parametrized
field theory
of \cite{kuchar1}, this representation is tailored to the
Heisenberg picture
quantization in which physical states are embedding
independent and
the action of spatial diffeomorphisms is trivial on the
field
variables:
\begin{equation}
(Q,T)\longrightarrow (Q,T\circ f).
\end{equation}

Presumably, there exists a representation in which 
the wavefunctionals and measure are separately invariant
under the 
action of spatial
diffeomorphisms that naturally
arise in the \Schrodinger picture quantization of
parametrized field theory
\cite{kuchar1}:
\begin{equation}
(Q,T)\longrightarrow (f^*Q,T\circ f) .
\end{equation}
We will explore this representation of the quantum field
theory
elsewhere.


\section*{5. Generalizations}

There are a number of ways one might try to generalize the
results presented in the previous sections.  Here we
briefly discuss partial results pertaining to such
generalizations; details will appear elsewhere.  The
generalizations that we consider include: inclusion 
of nonzero mass, massive and massless fields on flat
spacetimes diffeomorphic to $R\times R$, and
higher-dimensional generalizations of these models.

We begin by presenting a generic form for the Bogolubov
coefficient relevant for a discussion of unitary
implementability of dynamical evolution along an arbitrary
foliation.  We consider a free scalar field $\phii$
propagating on a flat $(n+1)$-dimensional spacetime $M$.  We
assume that 
$M\approx R\times\Sigma$, where either $\Sigma=
R^n$ 
or $\Sigma= {\bf T}^n$ (${\bf T}^n$ is the $n$-torus).  We
assume $\phii$ satisfies the Klein-Gordon equation
\begin{equation}
(\Box-m^2) \phii =0.
\label{eq:Klein-Gordon}
\end{equation} 
Let $T^\alpha$ and $x^i$ denote inertial coordinates on $M$
and arbitrary coordinates on $\Sigma$, respectively.  An
embedding $T\colon\Sigma\to M$ of a Cauchy surface is
represented by $n+1$ functions of $n$ variables:
\begin{equation}
T^\alpha=T^\alpha(x).
\end{equation}
The induced metric and future pointing unit normal of a
slice embedded by 
$T^\alpha(x)$ are denoted by $\gamma_{ij}$ and $n^\alpha$,
respectively.   Creation and annihilation operators 
$(a^\dagger_{\bf p}, a_{\bf p})$, are labeled by the wave
vector ${\bf p}$ for plane waves.  This vector takes on
discrete or continuous values when 
$\Sigma={\bf T}^n$ or $\Sigma=R^n$.
Dynamical evolution from an initial slice $T^\alpha_0(x)$
to a final slice 
$T^\alpha(x)$ can be viewed as
a symplectic transformation on the space
of solutions to 
(\ref{eq:Klein-Gordon}). 
Consequently, there is a corresponding Bogolubov
transformation
of the creation and annihilation operators.  If we choose
the initial embedding to be flat with Cartesian
coordinates, $T^\alpha(x)=(0,x^i)$, the mixing between
creation and annihilation operators is controlled by the
coefficients:
\begin{equation}
\beta_{{\bf k},{\bf p}}
={1\over\sqrt{\omega(k)\omega(p)}}
\int \left(\sqrt{\gamma}n^\alpha k_\alpha 
+\omega({\bf p})\right)
e^{-i({\bf p}\cdot {\bf x}+k_\alpha T^\alpha(x))}\,d^{n}x.
\label{eq:gen_bog}
\end{equation}
Here $\omega(k)=\sqrt{|{\bf k}|^2 + m^2}$ and
$k_\alpha=(-\omega(k),{\bf k})$.  We have dropped an
irrelevant
overall 
numerical factor in (\ref{eq:gen_bog}).

The Bogolubov coefficients (\ref{eq:gen_bog}) define an
operator $\beta$ on the one particle Hilbert space that
underlies the Fock space.  Unitary implementability of
dynamical evolution from $T_0^\alpha(x)$ to $T^\alpha(x)$
requires $\beta$ to be Hilbert-Schmidt.  We have seen that
this is so when 
$\Sigma=S^1$ and $m=0$ (there we had to also take account
of zero modes).  With compact spatial sections, the
Hilbert-Schmidt condition only involves the ultraviolet
behavior of
$\beta$, and one therefore expects that, for $\Sigma=S^1$,
$\beta$ is Hilbert-Schmidt
even when $m\neq0$.  This is indeed the case. We can prove
that dynamical evolution along arbitrary spacelike
foliations is unitarily implemented when $M=R\times
S^1$ for any value of the mass $m$.  When $M=R\times
R$ the massless case is rather similar to the
case studied in detail in the previous sections.  In
particular, we can show that the ultraviolet behavior 
of $\beta$ does not
spoil the Hilbert-Schmidt property provided the embeddings 
are asymptotically flat.  However, one
encounters an infrared divergence if one uses the usual
Schwartz space as the space of test functions.  We
expect that this case can nevertheless be handled with an
appropriate choice of test functions for operator valued
distributions representing the scalar field \cite{Glimm}. 
Likewise, we expect the 
operator $\beta$ for
a massive field on  $M=R\times R$
to be well-behaved in the infrared and ultraviolet for 
evolution involving asymptotically flat spacelike slices. 
Consequently, we conjecture that our results for a
massless, free, scalar field on $R\times S^1$ generalize to
any free field on a flat two-dimensional spacetime.  In
particular, we expect that dynamical evolution along
arbitrary spacelike foliations is unitarily implemented for
free fields on flat spacetimes $M=R\times S^1$ and along 
asymptotically flat spacelike foliations of
$M=R\times
R$.

The situation in higher dimensions is not nearly so simple
as it is for 
two-dimensional spacetimes.  It {\it is} possible to obtain
unitary 
evolution on the Fock space for free fields in higher
dimensions if 
one restricts attention to special classes of foliations. 
For example, 
dynamical evolution along a foliation obtained by dragging
an 
arbitrary spacelike slice along the integral curves of a
Killing 
vector field can be shown to be unitarily implementable.  
However, using the stationary phase
approximation, we have estimated (\ref{eq:gen_bog}) for the
case $\Sigma={\bf T}^n$ and found that $\beta$ is {\it not}
Hilbert-Schmidt for a generic embedding $T^\alpha(x)$. 
This means that dynamical evolution along arbitrary
spacelike foliations is {\it not} unitarily implemented in
the usual Poincare-invariant Fock representation for free
fields on flat spacetime.  A related difficulty is that the
smeared energy-momentum densities do not have the particle
number eigenstates ({\it e.g.}, the Fock vacuum) in their
domain (this point has already been noted in
\cite{Helfer}).  
This fact would explain the
divergent Schwinger terms that are encountered when
computing the algebra of energy-momentum tensors
\cite{torre}. We remark that an analogous situation 
arises in current
algebra \cite{rajeev}.  

It is an interesting open
question to find a Hilbert space quantization of  
free fields on flat
spacetime of dimension greater than two which yields the
correct physical results for dynamical evolution along
foliations by flat slices and which also allows for 
dynamical evolution along more
general foliations.  In particular, the standard apparatus
of Hilbert 
space and unitary time evolution does not seem adequate to
deal with 
quantization of parametrized field theory models of quantum 
gravity in spacetime dimensions greater than two.
It is well-known that analogous 
difficulties arise in the construction of quantum field
theories in 
curved spacetime, where generically there are no preferred
foliations 
available for the purposes of canonical quantization.  
In this case progress can be made by using 
algebraic methods of quantization (see {\it e.g.}
\cite{Waldbook}), 
and it is likely 
that such methods can be fruitfully applied to the class of
problems 
we are considering here.  
Thus, even in the simplest context 
of free fields in flat spacetime, our results suggest that
one is 
forced to abandon ``traditional'' approaches to
quantization of 
generally covariant theories in favor 
of the more flexible algebraic (or other) approaches.  

\section*{Acknowledgments}

\noindent 
The authors gratefully acknowledge helpful discussions of
this material with A. Ashtekar, S. Bose, G. Kang, K.
Kucha\v r, R. Nityananda,
J. Samuel, P. Sommers, 
and J. Whelan.  This work was supported in part by grants
PHY-9507719 (MV) and
PHY-9600616 (CGT) from the National Science Foundation.

\section*{Appendix}

In this appendix we show 
that the matrix $B^{(+)}_{mn}$ satisfies the Hilbert-Schmidt
condition (\ref{eq:HS}).

Since $T^+(x)$ is a diffeomorphism of the circle, it can be
used as a 
coordinate. Put $T^+(x)-T^+(0) = \theta$ and define $\chi$
to be 
the inverse function to 
$\theta$, that is,  $\chi (\theta ):= x$. Then 
\begin{equation}
B^{(+)}_{mn}= - {1\over2\pi}\sqrt{{n\over
m}}e^{inT^+(0)}\int_{0}^{2\pi}
e^{im\chi (\theta )+ in\theta }\, d\theta  \,  .
\end{equation}
For any $t\in [0,1]$, the function 
\begin{equation}
\chi_t(\theta ):= t\chi (\theta ) + (1-t) \theta
\end{equation}
is also a diffeomorphism. With $t= {m\over m+n}$,
\begin{equation}
B^{(+)}_{mn}= - {1\over2\pi}\sqrt{{n\over
m}}e^{inT^+(0)}\int_{0}^{2\pi}e^{
i(m+n)\chi_t(\theta )}\, d\theta  \,  .
\end{equation}
Put $\chi_t(\theta ) =y$ and denote the inverse function to
$\chi_t$ as 
$\varphi_t$.  Then 
\begin{equation}
B^{(+)}_{mn}= - {1\over2\pi}\sqrt{{n\over
m}}e^{inT^+(0)}\int_{0}^{2\pi}e^{
i(m+n)y} {d\varphi_t\over dy}\, dy  \,  .
\end{equation}
On integrating by parts $k$ times, 
\begin{equation}
B^{(+)}_{mn}= - {i^k\over2\pi}\sqrt{{n\over m}}(m+n)^{-k}
e^{inT^+(0)}\int_{0}^{2\pi}e^{
i(m+n)y} {d^{k+1}\varphi_t\over dy^{k+1}}\, dy  \,  ,
\end{equation}
which gives the estimate
\begin{equation}
|B^{(+)}_{mn}| \leq (n+m)^{-k}\sqrt{{n\over m}}
 sup\{|{d^{k+1}\varphi_t\over dy^{k+1}}| \,:\, 0\leq y\leq
2\pi, 0\leq t\leq 1\} .
\label{eq:bound}
\end{equation}
(Note that for sufficiently smooth embeddings 
$sup\{|{d^{k+1}\varphi_t\over dy^{k+1}}|\}$ exists).
Clearly (\ref{eq:bound})
 suffices to show that $B^{(+)}_{mn}$ is Hilbert-Schmidt.

Similar considerations, involving appropriate integrations
by parts,
suffice to show that $B^{(-)}_{mn}, \alpha_{(\pm)mn}$,
and that $\beta_{(\pm)mn}$ are Hilbert-Schmidt and that
$Z^{(\pm)}_n$
and $\zeta_{(\pm)n}$ are rapidly decreasing in $n$.


\begin{thebibliography}{999}
\bibitem{sign_error}
Note that this formula
corrects a
sign in \cite{kuchar1}.
\end{thebibliography}
\end{document}